\documentclass[sigconf]{acmart}

\AtBeginDocument{%
  \providecommand\BibTeX{{%
    \normalfont B\kern-0.5em{\scshape i\kern-0.25em b}\kern-0.8em\TeX}}}

\copyrightyear{2024}
\acmYear{2024}
\setcopyright{acmlicensed}
\acmConference[WWW '24] {Proceedings of the ACM Web Conference 2024}{May 13--17, 2024}{Singapore, Singapore.}
\acmBooktitle{Proceedings of the ACM Web Conference 2024 (WWW '24), May 13--17, 2024, Singapore, Singapore}
\acmPrice{15.00}
\acmISBN{979-8-4007-0171-9/24/05}
\acmDOI{10.1145/3589334.3645517}

\usepackage{algorithm}
\usepackage{algorithmic}
\usepackage{pdfpages}
\usepackage{multirow}
\usepackage[export]{adjustbox} 
\usepackage{bm}
\usepackage{amsmath}
\usepackage{array}
\usepackage{booktabs}
\usepackage{footmisc}

\usepackage{amsthm}
\usepackage{graphicx}
\usepackage{enumitem}

\theoremstyle{Definition}

\usepackage{subfigure}
\usepackage{hyperref}
\usepackage{url}
\usepackage{color}
\usepackage{xcolor}
\usepackage{colortbl}
\usepackage{xspace}
\usepackage{subcaption}
\newcommand{\model}{MacGNN\xspace}

\usepackage{amsmath,amsfonts,bm}

\def\MAC{{\widetilde{{\mathcal{C}}}}}
\def\MAE{{\widetilde{\bm{E}}}}

\def\MAR{{\widetilde{{\mathcal{R}}}}}
\def\MAn{{\widetilde{n}}}
\def\MAm{{\widetilde{m}}}

\def\eqref#1{equation~\ref{#1}}

\def\1{\bm{1}}

\def\mC{{\bm{C}}}

\DeclareMathAlphabet{\mathsfit}{\encodingdefault}{\sfdefault}{m}{sl}
\SetMathAlphabet{\mathsfit}{bold}{\encodingdefault}{\sfdefault}{bx}{n}

\def\gF{{\mathcal{F}}}
\def\gG{{\mathcal{G}}}

\def\gI{{\mathcal{I}}}

\def\gL{{\mathcal{L}}}

\def\gN{{\mathcal{N}}}
\def\gO{{\mathcal{O}}}

\def\gR{{\mathcal{R}}}
\def\gS{{\mathcal{S}}}
\def\gT{{\mathcal{T}}}
\def\gU{{\mathcal{U}}}

\definecolor{amaranth}{rgb}{0.9, 0.17, 0.31}
\definecolor{blue(pigment)}{rgb}{0.2, 0.2, 0.6}
\begin{document}

\title{Macro Graph Neural Networks for Online Billion-Scale Recommender Systems}

\author{Hao Chen}
\authornote{Both authors contributed equally to this research.}
\affiliation{%
  \institution{The Hong Kong Polytechnic University}
  \city{Hung Hom}
  \country{Hong Kong SAR}}
\email{sundaychenhao@gmail.com}

\author{Yuanchen Bei}
\authornotemark[1]
\affiliation{%
  \institution{Zhejiang University}
  \city{Hangzhou}
  \country{China}}
\email{yuanchenbei@zju.edu.cn}

\author{Qijie Shen}
\affiliation{%
  \institution{Alibaba Group}
  \city{Hangzhou}
  \country{China}}
\email{qijie.sqj@alibaba-inc.com}

\author{Yue Xu}
\affiliation{%
 \institution{Alibaba Group}
 \city{Hangzhou}
 \country{China}}
\email{yuexu.xy@foxmail.com}

\author{Sheng Zhou}
\affiliation{%
  \institution{Zhejiang University}
  \city{Hangzhou}
  \country{China}}
\email{zhousheng\_zju@zju.edu.cn}

\author{Wenbing Huang}
\affiliation{%
  \institution{Renmin University of China}
  \city{Beijing}
  \country{China}}
\email{hwenbing@126.com}

\author{Feiran Huang}
\authornote{Corresponding author.}
\affiliation{%
  \institution{Jinan University}
  \city{Guangzhou}
  \country{China}}
\email{huangfr@jnu.edu.cn}

\author{Senzhang Wang}
\affiliation{%
  \institution{Central South University}
  \city{Changsha}
  \country{China}}
\email{szwang@csu.edu.cn}

\author{Xiao Huang}
\affiliation{%
  \institution{The Hong Kong Polytechnic University}
  \city{Hung Hom}
  \country{Hong Kong SAR}}
\email{xiaohuang@comp.polyu.edu.hk}


\renewcommand{\shortauthors}{Hao Chen, et al.}

\begin{abstract}

Predicting Click-Through Rate (CTR) in billion-scale recommender systems poses a long-standing challenge for Graph Neural Networks (GNNs) due to the overwhelming computational complexity involved in aggregating billions of neighbors. To tackle this, GNN-based CTR models usually sample hundreds of neighbors out of the billions to facilitate efficient online recommendations. However, sampling only a small portion of neighbors results in a severe sampling bias and the failure to encompass the full spectrum of user or item behavioral patterns. To address this challenge, we name the conventional user-item recommendation graph as ``micro recommendation grap'' and introduce a revolutionizing \textbf{\underline{MA}cro Recommendation \underline{G}raph (MAG)} for billion-scale recommendations to reduce the neighbor count from billions to hundreds in the graph structure infrastructure. Specifically, We group micro nodes (users and items) with similar behavior patterns to form macro nodes and then MAG directly describes the relation between the user/item and the hundred of macro nodes rather than the billions of micro nodes. Subsequently, we introduce tailored \textbf{\underline{Mac}ro \underline{G}raph \underline{N}eural \underline{N}etworks (MacGNN)} to aggregate information on a macro level and revise the embeddings of macro nodes. MacGNN has already served Taobao's homepage feed for two months, providing recommendations for over one billion users. Extensive offline experiments on three public benchmark datasets and an industrial dataset present that MacGNN significantly outperforms twelve CTR baselines while remaining computationally efficient. Besides, online A/B tests confirm MacGNN's superiority in billion-scale recommender systems.

\end{abstract}



\maketitle

\section{Introduction}
Billion-scale recommender systems, with billions of users, items, and interactions, are prevalent in today's societies~\cite{ying2018graph,zhou2021contrastive,wu2022graph,zangerle2022evaluating}, such as YouTube~\cite{covington2016deep,zhou2010impact} and Taobao~\cite{pfadler2020billion,xu2023multi}. At the heart of these billion-scale recommender systems lies Click-Through Rate (CTR) prediction~\cite{ijcai2021p636,bei2023nrcgi}. Its goal is to predict, in real-time, whether a given user will click on a given item. However, due to efficiency requirements, while Graph Neural Networks (GNNs) have shown significant performance in collaborative filtering recommendation tasks~\cite{he2020lightgcn,chen2022gar}, they are not well-suited for CTR tasks. This is because performing graph neural networks over billion-scale neighbors leads to overwhelming computational complexity. It is crucial to develop appropriate graph neural networks capable of handling recommender systems with billions of users, items, and interactions.

Existing GNN models typically create the graph by linking users to their interacted (clicked) items~\cite{bei2023reinforcement,he2020lightgcn,wang2019neural}. In this scenario, if a user interacts with a highly popular item with billions of interactions, then the subgraph of that user will potentially have billions of 2-hop neighbors. To reduce computational complexity, PinSage~\cite{ying2018graph} randomly selects a fixed number of 1-hop and 2-hop neighbors for both users and items. GLSM~\cite{sun2022graph} and GMT~\cite{min2022neighbour} introduce importance-based and similarity-based scoring mechanisms to filter the most suitable hundreds of 1-hop and 2-hop neighbors. Besides, traditional CTR models introduce filtering strategies to accelerate the inferring process. DIN~\cite{zhou2018deep} and DIEN~\cite{zhou2019deep}, typically truncate a user's hundreds of recently interacted items. UBR4CTR~\cite{qin2020user} and SIM~\cite{pi2020search} further introduced search-based strategies to filter the most relevant items from the user's long historical behavior. However, traditional CTR models fail to consider the subgraph of items or the 2-hop neighbors of users.

\begin{figure}
  \centering
  \subfigure[Neighborhood number distribution of micro graphs.]{
    \centering
    \includegraphics[width=1\linewidth]{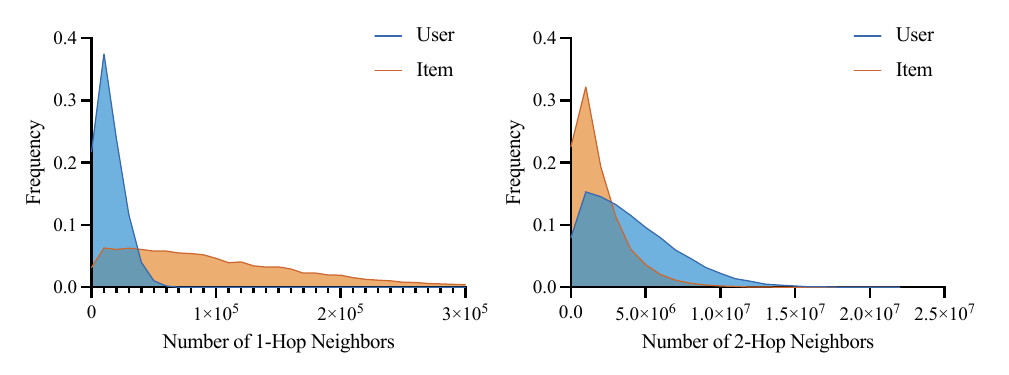}

    \label{fig:intro-micro}
  }

  \subfigure[Neighborhood number distribution of macro graphs.]{
    \centering
    \includegraphics[width=1\linewidth]{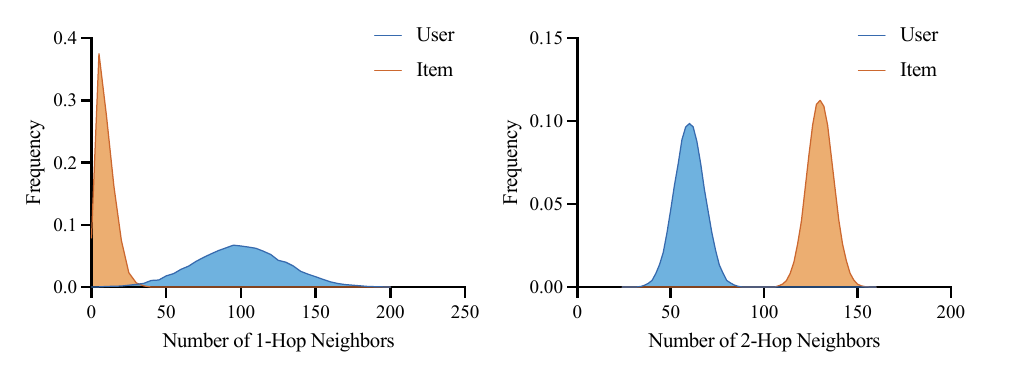}
    
    \label{fig:intro-macro}
  }
    \vspace{-1em}
  \caption{Illustration of neighbor number distributions in micro and macro user-item clicking interaction graphs within Taobao's billion-scale recommender system.}
  \vspace{-1em}
  \label{fig:vertical_subfigures}
\end{figure}

Though the above strategies can reduce the neighbor size for GNNs, these approaches still face the following limitations in billion-scale recommender systems.
\\
\textbf{1. Severe Sampling Bias:} In Figure~\autoref{fig:intro-micro}, we illustrate the distribution of neighbor numbers in the user-item clicking interaction graph within a real-world shopping platform. Both users and items exhibit a substantial number of 1-hop and 2-hop neighbors. Sampling only a few hundred neighbors can only cover about 5\% of user 1-hop neighbors and 0.2\% of item 1-hop neighbors. Sampling such small portions cannot accurately represent the entire spectrum of neighbors and may lead to severe sampling bias.
\\
\textbf{2. Unfitted Users/Items Sampling:}
   As shown in Figure~\autoref{fig:intro-micro}, users exhibit vastly different number distributions compared to items. For example, users have significantly more 2-hop neighbors and significantly fewer 1-hop neighbors than items. It is inappropriate to sample users and items using the same approach.
\\
\textbf{3. Ambiguous Neighbor Counts:}
   The sampled neighbors do not accurately represent the true number of interactions prior to the sampling process for users and items. For instance, a user with hundreds of historical interactions will yield the same sample size as another user with millions of historical interactions.

The main problem behind the mentioned issues arises from relying on sampling strategies to decrease the size of neighbors. Instead, it's more promising to boost the expressive capacity of graph nodes and significantly reduce the neighbor size by grouping nodes into macro nodes. This grouping approach allows models to overcome the inherent limitations of sampling strategies by eliminating the need for the sampling process entirely. However, actualizing a grouping strategy for recommendation graphs introduces the following challenges.
\\
\textbf{1. Grouping Strategy:} Identifying an optimal grouping strategy for user and item nodes into macro nodes is non-trivial, as it demands a careful balance between reducing complexity and maintaining the integrity of original behavioral patterns. 
\\
\textbf{2. Subgraph Definition:} Constructing edges between macro nodes is complex due to the necessity of representing aggregated interactions between their constituent user/item nodes accurately. Additionally, defining the subgraph for a given user/item using macro nodes demands innovative approaches.
\\
\textbf{3. Recommending with Macro Nodes:} Each macro node represents a group of user/item nodes, and the edge between two macro nodes signifies the connections between two groups of nodes. It is challenging to extract the behavioral pattern of a user/item node based on its newly constructed macro-node subgraphs.


\begin{figure}[tbp]
     \centering
     \includegraphics[width=0.95\linewidth, trim=0cm 0cm 0cm 0cm,clip]{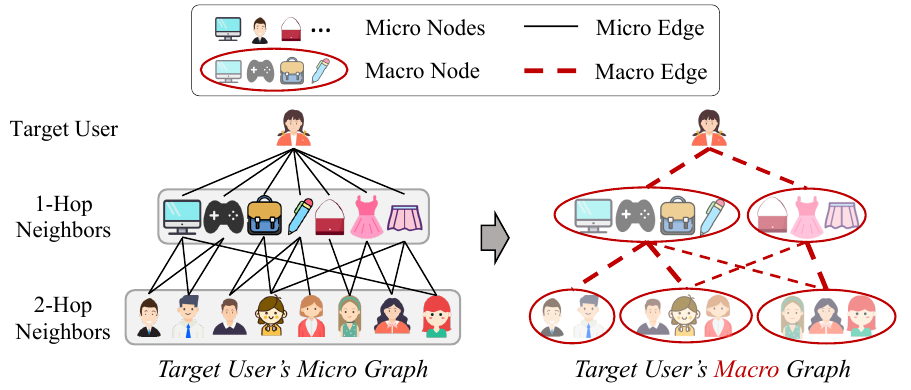}
     \vspace{-0.5em}
     \caption{Sketch map of the construction of the macro graph.}
     \vspace{-0.8em}
     \label{fig:macrograph}
\end{figure}

By addressing the three challenges mentioned above, we propose a more suitable \textbf{\underline{MA}cro Recommendation \underline{G}raph (MAG)} for billion-scale recommendations. MAG groups user/item nodes based on similar behaviors to create macro nodes, as illustrated in \autoref{fig:macrograph}. This grouping reduces the number of neighbors from billions to hundreds. As depicted in Figure \autoref{fig:intro-macro}, MAG now only consists of hundreds of 1-hop and 2-hop neighbors. This reduction allows billion-scale recommender systems to alleviate the adverse consequences of sampling only a small portion of neighbors. To achieve this, we introduce tailored \textbf{\underline{Mac}ro \underline{G}raph \underline{N}eural \underline{N}etworks (MacGNN)} to aggregate the macro information for the target user/item with our specially designed MAG, facilitating accurate and efficient click-through rate prediction for online billion-scale recommender systems. Our paper's primary contributions are summarized as follows:
\begin{itemize}[leftmargin=*]
    \item We create a customized macro recommendation graph, which involves constructing the macro node, macro edge, and macro subgraph. This helps reduce the neighbor size from billions to hundreds, making it easier for GNNs to operate in online billion-scale recommender systems.
    \item We propose a novel macro-scale recommendation paradigm known as the Macro Graph Neural Network (MacGNN). This framework efficiently aggregates macro-graph information and updates macro-node embeddings to enable online click-through rate prediction for billion-scale recommender systems.\footnote{Source code is available at \url{https://github.com/YuanchenBei/MacGNN}.}
    \item MacGNN has been serving a major shopping platform for two months, offering recommendations to more than one billion users. Additionally, we introduce our online implementation to enable online updates of macro nodes and macro edges.
    \item Extensive offline experiments conducted on three public benchmark datasets and a billion-scale industrial dataset demonstrate that MacGNN outperforms twelve state-of-the-art CTR baselines while maintaining competitive efficiency. Furthermore, online A/B tests have confirmed the superiority of MacGNN in real-world billion-scale recommender systems.
\end{itemize}

\section{Preliminaries}
In this section, we first present the basic notations in CTR prediction. Then, we present the concept of micro nodes, micro edges, and micro recommendation graphs for recommender systems. Finally, we introduce the definition of our macro recommendation graph.

\textit{\textbf{CTR Prediction.}} Supposed the set of users and items as $\gU=\{u_{1}, ..., u_{n}\}$, and $\gI=\{i_{1}, ..., i_{m}\}$, respectively, where $|\gU|=n$ and $|\gI|=m$ denotes the number of users and items. In real-world recommender systems, CTR models correspond to a click or not problem. When item $i$ is exposed to user $u$, user $u$ will have two reflections: (i) having a positive behavior toward the item $i$ such as click or purchase, or (ii) having a negative behavior toward the item $i$ such as neglect or dislike. Thus, given the target user-item pair as $(u,i)$, the corresponding interaction $y_{ui}$ can be present as:
\begin{equation} \label{eq:ctr}
y_{ui} = 
\begin{cases} 
    1, & \text{if } u \text{ exhibits positive behavior towards } i; \\
    0, & \text{if } u \text{ exhibits negative behavior towards } i.
\end{cases}
\end{equation}

Given a target user-item pair $(u, i)$, the CTR prediction task is to predict the target user $u$'s positive behavior probability $\hat{y}_{ui}$ on target item $i$. In form, the aim of a CTR model is to learn an accurate prediction function $\gF(\cdot)$, namely the predicted clicking probability $\hat{y}_{ui}=\gF(u, i)$, to minimize the difference from $\hat{y}_{ui}$ to $y_{ui}$.

\textit{\textbf{Micro Node.}} Starting with several popular works~\cite{ying2018graph,wang2019neural,he2020lightgcn}, GNN-based recommendation models usually connect users with their interacted~(e.g. clicked or purchased) items. Under this setting, users and items are treated as micro nodes. Specifically, each user $u$ and item $i$ is associated with a trainable embedding $\bm{E}_{u}\in\mathbb{R}^{d}$ and $\bm{E}_{i}\in\mathbb{R}^{d}$, where $d$ is the embedding dimension size. 

\textit{\textbf{Micro Edge.}} As stated in Eq.~(\ref{eq:ctr}), the user-item behaviors actually provide the most raw material for edges. Given the micro user-item interaction matrix $\gR \in \mathbb{R}^{|\gU|\times|\gI|}$, where $|\gR|$ is the total number of interactions. Each element $r_{ui}\in\gR$ reflects whether users $u$ have a positive interaction with item $i$, namely $r_{ui} = y_{ui}$.

\textit{\textbf{\underline{MI}cro Recommendation \underline{G}raph (MIG).}} After defining the micro nodes and micro edges, the MIG can be represented as $\gG = (\gU,\gI,\gR)$.
For user interest models, the user behavior sequences can be given as the first-order neighbor of the user $u$ as $\gN_{u}^{(1)}$, where $\gN_{u}^{(k)}$ denotes the $k^{th}$-hop neighbors of user $u$.

According to the definition of MIG, when GNNs predict the CTR of a given user-item pair $(u,i)$, GNNs first construct the micro subgraph of the target user/item and then extract the embeddings according to MIG. When the graph size grows to a billion-scale, the subgraph may contain billions of micro nodes, which means only loading the embeddings of the subgraph is difficult to accomplish.

\textit{\textbf{\underline{MA}cro Recommendation \underline{G}raph (MAG).}}
Our proposed MAG can be defined as $\widetilde{\gG} = (\widetilde{\gU}, \widetilde{\gI}, \widetilde{\gR})$, where $\widetilde{\gU}$, $\widetilde{\gI}$, and $\widetilde{\gR}$ are the macro user nodes, macro item nodes, and macro edges respectively, and $\widetilde{\gN}_v^{(k)}$ represents the $k^{th}$-hop macro neighbors of node $v$. 
Specifically, each macro node $v$ is associated with a trainable embedding $\MAE_{v}\in\mathbb{R}^{d}$. With MAG, MacGNN only needs to aggregate hundreds of macro nodes, significantly reducing computational complexity.

\section{Methodology}

In this section, we first formally introduce the concept of Macro Recommendation Graphs and introduce how to design macro nodes and macro edges. Then we present the macro graph neural network for CTR prediction.
Finally, we illustrate the implementation architecture of our real-world billion-scale recommender system.

\subsection{Macro Recommendation Graph (MAG)}

\subsubsection{\textbf{Constructing Macro Nodes}}
As presented in the preliminaries, MIG records the detailed micro node and micro edge for each user and item. Then, given any user or item, the GNNs have to access the embeddings of each hop of micro nodes to infer the behavior pattern of the given user or item, which is computationally inconvenient and raises responsible delays. Motivated by this, MAG presents the behavior pattern within macro nodes rather than listing all the micro nodes and utilizes the GNNs to extract the behavior pattern from detailed micro nodes. 

Intuitively, the macro nodes are designed to represent the behavior pattern of a set of micro nodes, while all the micro nodes inside share similar behavior patterns.
Thus, we conduct the behavior pattern grouping to map the micro nodes into specific macro behavior nodes, with the objection of minimizing the behavioral pattern gap between micro nodes assigned to the same macro node~\cite{hamerly2003learning}.

Specifically, given the micro user-item interaction matrix $\gR\in \mathbb{R}^{|\gU|\times|\gI|}$, for a given user/item micro node $v$, we first obtain its behavior embedding $\bm{b}_{v}$ as follows:
\begin{equation}
    \bm{b}_{v} = \frac{[\bm{R}]_{v}}{\parallel[\bm{R}]_{v}\parallel_{2}} = \frac{\bm{r}_{v}}{\parallel\bm{r}_{v}\parallel_{2}},\qquad \bm{R} = 
    \begin{cases}
    \gR,& v \in \gU;\\
    \gR^\top,& v \in \gI .
    \end{cases}
\end{equation}
where $||\cdot||^{2}_{2}$ is the $L_2$ norm. 
Then, to obtain each macro node $\mC_{k}$, we conduct the behavior pattern grouping based on behavior embeddings of micro nodes. Specifically, we first randomly initialize $K$ macro centroids $\{\bm{\mu}_{1}, ..., \bm{\mu}_{k}, ..., \bm{\mu}_{K}\}$, 
where $\bm{\mu}_{k}\in \mathbb{R}^{d}$ is the centroids of macro node $\mC_{k}$, $K \ll n \ and \ m$ is the hyperparameter set as the macro node number, and we denote $K$ for macro user node and macro item node is $\MAn$ and $\MAm$, respectively. Then, we explore and assign micro nodes to the appropriate macro node based on their behavior patterns, and update the centroid of macro nodes iteratively. The process can be expressed as:
\begin{equation}
    \bm{\mu}_{k} = \frac{1}{|\mC_{k}|}\sum_{x_{v}=k, \bm{b}_v \in \mC_{k}} \bm{b}_{v},
\end{equation}
where $|\mC_{k}|$ is the number of micro nodes within $\mC_{k}$, and $x_{v}$ is the macro node index that $v$ is assigned to.
Further, the optimization objection of the behavior pattern grouping is:
\begin{equation}
\begin{aligned}
    \min_{x_1, ..., x_{m+n} \atop \bm{\mu}_{1}, .., \bm{\mu}_{K}} &J\left(x_1, ..., x_{m+n}; \bm{\mu}_{1}, .., \bm{\mu}_{K} \right)
    \\ \overset{\bigtriangleup}{=} & \sum_{k=1}^{K}\sum_{x_{v}=k, \bm{b}_{v}\in \mC_{k}}\sqrt{(\bm{b}_{v}-\bm{\mu}_{k})(\bm{b}_{v}-\bm{\mu}_{k})^\top}.
\end{aligned}
\end{equation}
where $J$ is the objection function of behavior pattern grouping.
As shown in~\autoref{fig:macrograph}, the micro nodes with similar behavior patterns will be composed of a macro node.
Note that each macro node $v$ will also be assigned a trainable embedding $\MAE_{v}\in \mathbb{R}^{d}$.


\subsubsection{\textbf{Organizing Macro Edges}}
Macro edges depict relationships between two macro nodes within a specific user/item subgraph, signifying the behavioral patterns within that subgraph. It's important to note that macro edges have a distinct design compared to micro edges. The micro edges present connections between fixed micro user nodes and micro item nodes. Since micro nodes remain constant, the micro edges are also fixed. In contrast, macro edges capture the connection strength between two macro nodes in a subgraph, which is tailored to each user and item subgraph.


In~\autoref{fig:macrograph}, the user $v$ is depicted as having two 1-hop macro nodes, each with macro edge weights of 4 and 3, respectively. Moving to the second hop, the user extends to three macro nodes, and these macro edges represent the connections between the 1-hop macro nodes and the 2-hop macro node. Formally, we use $\MAC = \{\mC_1, \mC_2, \ldots, \mC_{\MAn+\MAm}\}$ to represent the entire set of macro nodes in the MAG. We employ $\MAR^{(k)}_{v;p,q}$ to denote the macro edge for any user/item node $v$ with its $k^{th}$-hop neighbors, where $\mC_{v;p}^{(k-1)}$ represents the macro node in $(k-1)^{th}$-hop macro neighbors $\widetilde{\gN}_v^{(k-1)}$, and $\mC_{v:q}^{(k)}$ represents the macro node in $k^{th}$-hop macro neighbors $\widetilde{\gN}_v^{(k)}$. Thus the weight of macro edges can be computed as:
\begin{equation}
    \label{eq:macro_edge}
    \MAR^{(k)}_{v;p,q} = \sum_{a\in \mC_{v;p}^{(k-1)},b\in\mC_{v:q}^{(k)}} r_{ab},
\end{equation}
where $\mC_{v;p}^{(k-1)} = \mC_{v;p} \cap \gN_v^{(k-1)}$ represents the macro nodes related to node $v$ within its $(k-1)^{th}$-hop neighbors and $\mC_{v;q}^{(k)} = \mC_{v;q} \cap \gN_v^{(k)}$ represents the macro nodes related to node $v$ within its $k^{th}$-hop neighbors. In~\textsection~\ref{sec:online}, we will introduce how to get online updating macro edges on billion-scale recommender systems.
Finally, after transforming micro recommendation graphs into macro recommendation graphs, MAGs have significantly fewer nodes and edges by extracting behavior patterns explicitly into macro nodes. 
\begin{figure}[tbp]
    \centering
    \includegraphics[width=\linewidth, trim=0cm 0cm 0cm 0cm,clip]{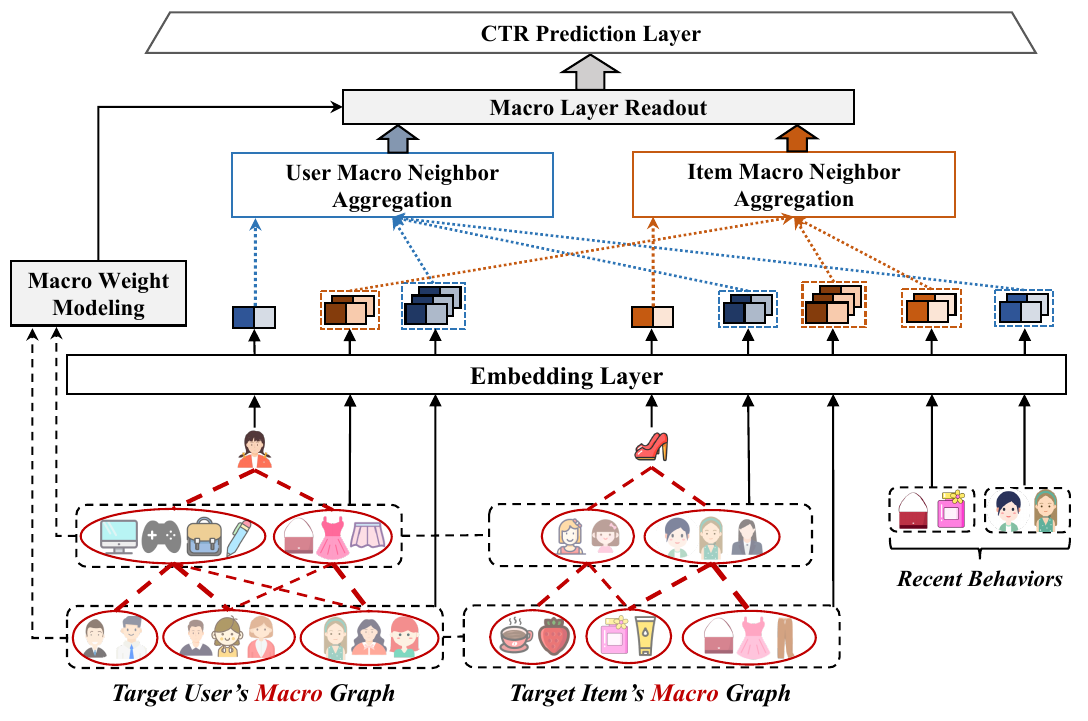}
    \caption{The model architecture of the proposed \model.}
    \vspace{-1em}
    \label{fig:framework}
\end{figure}

\subsection{Macro Graph Neural Network}

\subsubsection{\textbf{Macro Weight Modeling}}
The overall framework of our proposed \model is shown in~\autoref{fig:framework}.
To better identify the target user/item preferences over a certain macro node, we design the macro weight modeling for macro neighbors according to the weights of connected macro edges.

In order to avoid the excessive gap between the macro edge weights of hot nodes and cold nodes and conduct modeling flexibly, we equip the macro weight modeling with logarithmic smoothing and temperature-based softmax activation.
Formally, take the target user/item $v$ as an example, given a macro node $q$ in its $k^{th}$-hop neighborhood, the macro weight $w_{v;q}^{(k)}$ of $q$ toward the target user/item $v$ is calculated as:
\begin{equation}
    s_{v;q}^{(k)} = \log\left(\sum_{p\in \widetilde{\gN}_v^{(k-1)}}\MAR^{(k)}_{v;p,q}+1\right),\quad w_{v;q}^{(k)} = \frac{exp\left(s_{v;q}^{(k)}/\tau\right)}{\sum_{j\in \widetilde{\gN}_v^{(k)}}exp\left(s_{v;j}^{(k)}/\tau\right)},
\end{equation}
where $\tau$ is a temperature coefficient hyper-parameter~\cite{chen2023adap}. These modeled weights represent the importance of these macro neighboring nodes in the target user/item's historical interactions.


\subsubsection{\textbf{Marco Neighbor Aggregation \& Layer Readout}}
To mine the macro relationships effectively and efficiently, we first design a macro neighbor aggregation architecture rather than a time-consuming recursive graph convolution. Then, we propose the macro layer readout to aggregate the macro information of the target user and item.

\textit{\textbf{Macro Neighbor Aggregation.}} Due to the different semantics of users and items, we utilized two separate macro neighbor aggregation modules without parameter sharing for user-type macro nodes and item-type macro nodes, respectively.

For user-type target nodes and their $k^{th}$-hop user-type macro neighbors, the aggregation function $MNA_{u}$ can be defined as:
\begin{equation}
\begin{aligned}
    MNA_{u}&(u, p\in \widetilde{\gN}_u^{(k)}, \bm{E}_{u}, \MAE_{p}, \widetilde{\gN}_u^{(k)}; \bm{Q}_{u}, \bm{K}_{u}, \bm{V}_{u}) 
    \\ \overset{\bigtriangleup}{=} & \sum_{p\in \widetilde{\gN}_u^{(k)}} \sigma\left(\langle \bm{Q}_{u}\cdot \MAE_{p}, \bm{K}_{u}\cdot \bm{E}_{u} \rangle \right) \cdot \bm{V}_{u} \cdot \MAE_{p},
\end{aligned}
\end{equation}
where $\bm{Q}_{u}, \bm{K}_{u}, \bm{V}_{u} \in \mathbb{R}^{d\times d'}$ are trainable self-attention matrics for user-type nodes, $\langle\cdot\rangle$ is the inner product function, and $\sigma(\cdot)$ is the softmax activation function.
Specifically, given the given target user $u$ and a user-type macro node $p$ in its $k^{th}$-hop neighborhood (such as the node in target user $u$'s 2-hop macro neighborhood and target item $i$'s 1-hop macro neighborhood), the process is expressed as:

\begin{equation}
    \alpha_{u,p} = \frac{exp\left((\bm{Q}_{u} \cdot \MAE_{p})(\bm{K}_{u} \cdot \bm{E}_{u})^\top/\sqrt{d}\right)}{\sum_{j\in \widetilde{\gN}_u^{(k)}}exp\left((\bm{Q}_{u} \cdot \MAE_{j})(\bm{K}_{u} \cdot \bm{E}_{u})^\top/\sqrt{d}\right)},
\end{equation}
\begin{equation}
    \widetilde{\bm{Z}}_{u,p} = \alpha_{u,p}\cdot(\bm{V}_{u}\cdot \MAE_{p}),
\end{equation}
where $\widetilde{\bm{Z}}_{u,p}$ is the aggregated macro embedding.
Similarly, for the item-type target node $i$ and its macro item-type neighbor $p$ in the $k^{th}$-hop neighborhood, the aggregation function $MNA_{i}$ to obtain the aggregated macro embedding $\widetilde{\bm{Z}}_{i,q}$ can be derived in similar ways using separating parameters as:
\begin{equation}
\begin{aligned}
     MNA_{i}&(i, q\in \widetilde{\gN}_i^{(k)}, \bm{E}_{i}, \MAE_{q}, \widetilde{\gN}_i^{(k)}; \bm{Q}_{i}, \bm{K}_{i}, \bm{V}_{i}) 
    \\ \overset{\bigtriangleup}{=} & \sum_{q\in \widetilde{\gN}_i^{(k)}} \sigma\left(\langle \bm{Q}_{i}\cdot \MAE_{q}, \bm{K}_{i}\cdot \bm{E}_{i} \rangle \right) \cdot \bm{V}_{i} \cdot \MAE_{q},
\end{aligned}
\end{equation}
where $\bm{Q}_{i}, \bm{K}_{i}, \bm{V}_{i} \in \mathbb{R}^{d\times d'}$ are trainable self-attention matrics for item-type nodes.


\textit{\textbf{Macro Layer Readout.}}
With the co-consideration of macro weight modeling and macro neighbor aggregation, we can measure the importance of the specific neighboring macro node from different perspectives.
Thus, the representation of a specific-hop macro neighborhood of the target user/item node can be obtained by the following layer readout:
\begin{equation}
    \bm{E}_{u}^{(l_{u})} = \sum_{j\in \widetilde{\gN}_u^{(l_{u})}} w_{u,j}\cdot \widetilde{\bm{Z}}_{u,j},\quad \bm{E}_{i}^{(l_{i})} = \sum_{j\in \widetilde{\gN}_i^{(l_{i})}} w_{i,j}\cdot \widetilde{\bm{Z}}_{i,j},
\end{equation}
where $\bm{E}_{u}^{(l_{u})}$ and $\bm{E}_{i}^{(l_{i})}$ denote the $l_{u}$-hop/$l_{i}$-hop readout representation of target user/item, respectively.

\subsubsection{\textbf{Recent Behavior Modeling}}
The above macro modeling takes into account the general and stable behavioral characteristics of the target node. Leveraging the learned knowledge at such a macro level, we further consider the information of recent behavior to better extract users' changing short-term interests and the evolving interaction patterns of items~\cite{deng2018ad,sun2022graph}.

Formally, for the target user $u$ and target item $i$, the few most recently interacted neighbor sequence $\bm{RS}_{u}$ and $\bm{RS}_{i}$ are utilized and their embeddings are co-trained with the macro nodes in the above aggregation functions, respectively.
\begin{equation}
\begin{aligned}
    \bm{Z}_{u,rs_{p}} &= MNA_{i}(i, rs_{p} \in \bm{RS}_{u}, \bm{E}_{i}, \bm{E}_{rs_{p}}, \bm{RS}_{u}; \bm{Q}_{i}, \bm{K}_{i}, \bm{V}_{i}),\\
    \bm{Z}_{i,rs_{q}} &= MNA_{u}(u, rs_{q} \in \bm{RS}_{i}, \bm{E}_{u}, \bm{E}_{rs_{q}}, \bm{RS}_{i}; \bm{Q}_{u}, \bm{K}_{u}, \bm{V}_{u}),
\end{aligned}
\end{equation}
\begin{equation}
    \bm{E}_{i}^{rs} = \sum_{rs_{q}\in \bm{RS}_{i}}\bm{Z}_{i,rs_{q}},\quad \bm{E}_{u}^{rs} = \sum_{rs_{p}\in \bm{RS}_{u}}\bm{Z}_{u,rs_{p}},
\end{equation}
where $\bm{E}_{u}^{rs}$ and $\bm{E}_{i}^{rs}$ are the representation of the few macro node sequence $\bm{RS}_{u}$ and $\bm{RS}_{i}$. The sequence length of the few recent behaviors for auxiliary training is set to 20.
Note that the number of recent nodes for modeling is much smaller than the hundreds of sequence lengths in the advanced interest models~\cite{zhou2018deep,zhou2019deep}.

\subsubsection{\textbf{CTR Prediction Layer}}
With the obtained informative representations, we utilize them for the final CTR prediction for the target user $u$ and target item $i$ as the following calculation:
\begin{equation}
    \hat{y}_{u,i} = MLP\left((\Arrowvert_{l_u}^{K}\bm{E}_{u}^{(l_u)})\parallel(\Arrowvert_{l_i}^{K}\bm{E}_{i}^{(l_i)})\parallel \bm{E}_{u}^{rs} \parallel \bm{E}_{i}^{rs} \parallel \bm{E}_{u} \parallel \bm{E}_{i}\right),
\end{equation}
where the architecture and parameter settings of the MLP are the same as previous works~\cite{zhou2018deep,zhou2019deep}.

To train and optimize the model parameters, we apply the binary cross-entropy loss as the model objective function.
Formally, for each user-item pair $(u,i)$ in training set $\gT\gS$, the adopted objective function can be expressed as:
\begin{equation}
    \gL_{bce} = -\frac{1}{|\gT\gS|} \sum_{(u,i)\in \gT\gS} y_{u,i} \log(\hat{y}_{u,i}) + (1-y_{u,i}) \log(1-\hat{y}_{u,i}),
\end{equation}
where $\hat{y}_{u,i}$ is the predicted CTR and $y_{u,i}$ is the ground-truth label.
Then, the overall objective function of \model is as follows:
\begin{equation}
    \gL=\gL_{bce} + \lambda \cdot \left\lVert \bm{\theta} \right\rVert_{2}^{2},
\end{equation}
where $\lambda \cdot \left\lVert \bm{\theta} \right\rVert_{2}^{2}$ denotes the $L_2$ regularization to avoid over-fitting. 

\subsection{Online Implementation}
\label{sec:online}
\begin{figure}[tb]
    \centering
    \includegraphics[width=\linewidth, trim=0cm 0cm 0cm 0cm,clip]{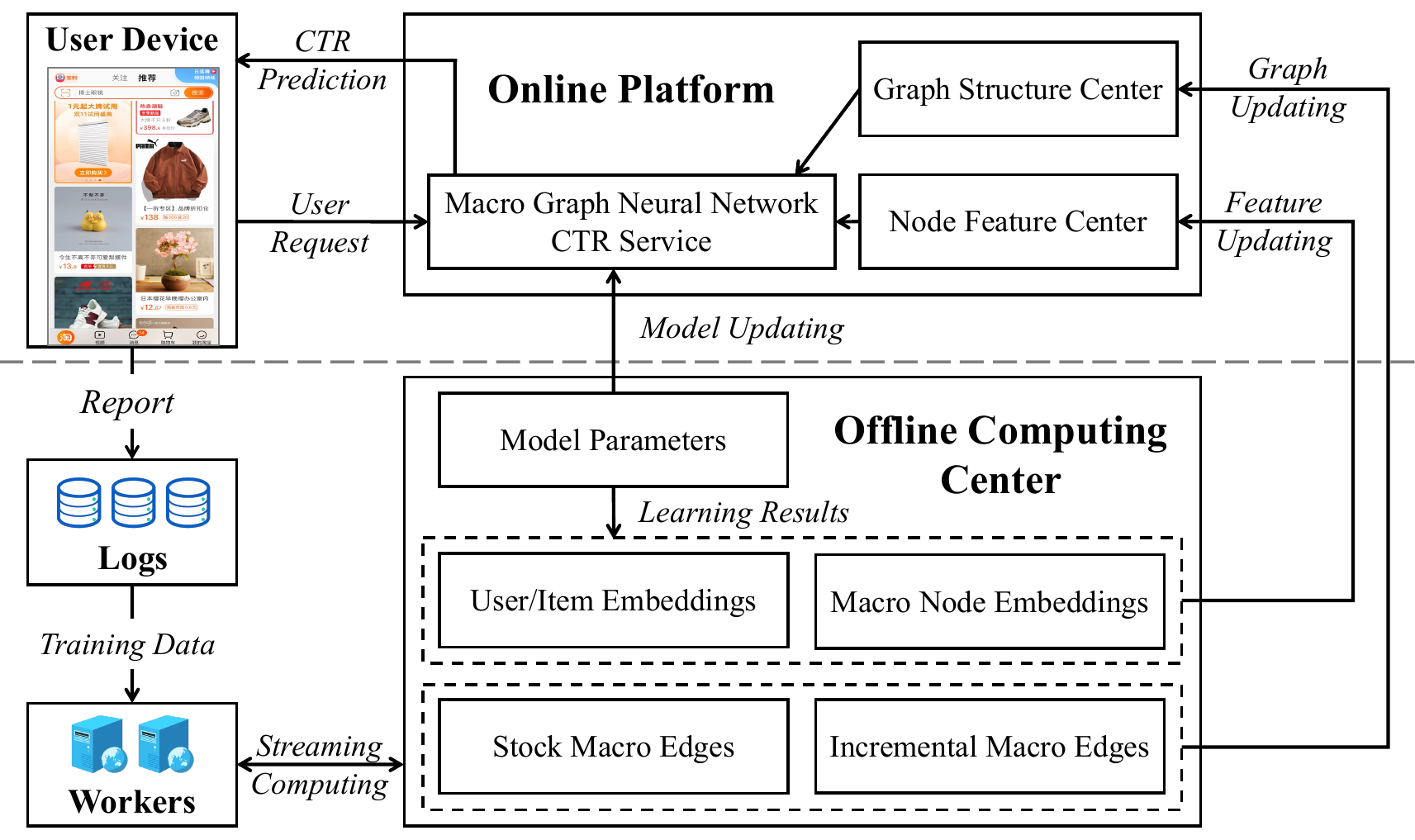}
    \caption{The system architecture for online deployment.}
    \vspace{-0.8em}
    \label{fig:onlineArch}
\end{figure}
In this section, we present the online deployment of MacGNN on a leading e-commerce platform's homepage. MacGNN has provided stable and precise recommendations to over 1 billion users and 2 billion items, analyzing more than 12 trillion interactions since August 2023.

The core architecture to implement the proposed MacGNN model is presented in Fig.~\ref{fig:onlineArch}, including the workflow of both offline computing and online serving. Offline computing can compute the necessary embeddings and graph structures without affecting the online service.
Specifically, offline computing is based on a distributed machine learning platform, which loads log data to train the model parameters and embeddings. Then the learned user/item embedding and the macro node embedding are uploaded to the graph feature center for online serving. 

Another job of offline computing is the graph structure updates. For example, during shopping events like Black Friday or Singles' Day, certain popular items can receive billions of clicks within seconds. In such scenarios, we employ two modules to facilitate graph structure updates. The stock micro edges are computed offline on a daily basis (or even hourly if necessary). Meanwhile, the incremental micro edges store the micro edges generated in real-time. Since the macro edge weights (Eq. (\ref{eq:macro_edge})) are defined through summation, the complete micro edge weights can be computed by adding the stock macro edge weights and the incremental macro edge weights.

With the help of offline computing, during the online inferring process, MacGNN can directly get the macro edges through the graph structure center and get the macro node embeddings through the graph feature center. Since MacGNN only considers the macro node, we can give the upper bound of the related node number as $\gO(({\MAn}+{\MAm}))$. On the contrary, the expected related node number of traditional micro GNNs can be given as $\gO(\frac{|\gR|^2}{m\times n})$. 
Specifically, we construct 200 macro nodes for users and 300 macro nodes for items. Then the micro GNNs will consider about 6 million times more nodes of the MacGNN if micro GNNs consider all the micro nodes in the billion-scale recommender system.

\section{Experiments}

In this section, we conduct comprehensive experiments on both offline datasets and real-world online recommendation systems, aiming to answer the following research questions.
\textbf{RQ1:} How does \model perform compared to state-of-the-art models?
\textbf{RQ2:} How efficient is the proposed \model?
\textbf{RQ3:} What is the effect of different components in \model?
\textbf{RQ4:} How does \model perform on billion-scale real-world recommendation platforms?

\subsection{Experimental Setup}
\subsubsection{Datasets}
We conduct comprehensive experiments on three widely used benchmark datasets \textbf{MovieLens}~\cite{harper2015movielens}, \textbf{Electronics}~\cite{mcauley2015image}, and \textbf{Kuaishou}~\cite{gao2022kuairec}, and one \textbf{billion-scale industrial dataset from Alibaba}, one of the biggest shopping platforms in China, to verify the effectiveness of \model. The statistics of these datasets are shown in Table \ref{tab:stats}.
The detailed description of these datasets is illustrated in Appendix \ref{sec:app_data}.

\subsubsection{Competitors}
To evaluate the effectiveness of \model, we compare it with twelve representative state-of-the-art CTR prediction models into three main groups.
(i) \textit{Feature Interaction-based Methods}: \textbf{Wide\&Deep}~\cite{cheng2016wide}, \textbf{DeepFM}~\cite{guo2017deepfm}, \textbf{AFM}~\cite{xiao2017attentional}, and \textbf{NFM}~\cite{he2017nfm}.
(ii) \textit{User Interest-based Methods}:
\textbf{DIN}~\cite{zhou2018deep}, \textbf{DIEN}~\cite{zhou2019deep}, \textbf{UBR4CTR}~\cite{qin2020user}, and \textbf{SIM}~\cite{pi2020search}.
(iii) \textit{Graph-based Methods}:
\textbf{PinSage}~\cite{ying2018graph},
\textbf{LightGCN}~\cite{he2020lightgcn}, 
\textbf{GLSM}~\cite{sun2022graph}, and \textbf{GMT}~\cite{min2022neighbour}.
We leave the details of these baseline models in Appendix \ref{sec:app_baseline}.


\subsubsection{Hyperparameter Setting}
For all models, the embedding size is fixed to 10 and the embedding parameters are initialized with the Xavier method~\cite{glorot2010understanding} for fair comparison. 
Shapes of the final MLP for all models are set to [$200, 80, 2$] as previous works~\cite{zhou2018deep,zhou2019deep}.
The learning rate of \model is searched from \{$1 \times 10^{-2},\ 5 \times 10^{-3},\ 1 \times 10^{-3}$\}, the regularization term $\lambda$ is searched from \{$1 \times 10^{-4},\ 5 \times 10^{-5},\ 1 \times 10^{-5}$\}. The batch size is set to 1024 for all models and the Adam optimizer is used~\cite{kingma2014adam}.

\subsubsection{Evaluation Metrics}
We evaluate the models with three widely-adopted CTR prediction metrics including AUC~\cite{fawcett2006introduction}, GAUC~\cite{zhou2018deep}, and Logloss~\cite{chen2016logloss}. The \textit{higher} AUC and GAUC values indicate higher CTR prediction performance, and the \textit{lower} Logloss value indicates higher CTR prediction performance.
Note that we run all the experiments \textit{five} times with different random seeds and report the average results with standard deviation to prevent extreme cases.

\begin{table}[tbp]
  \centering
  \small
  \caption{Statistics of the experimental datasets.}
  \vspace{-0.25em}
  \resizebox{0.985\linewidth}{!}{
    \begin{tabular}{c|c|c|c|c}
    \toprule
    Dataset & \# Users & \# Items & \# Interactions & \# Categories \\
    \midrule
    MovieLens & 71,567 & 10,681 & 10,000,054 & 21 \\
    Electronics & 192,403 & 63,001 & 1,689,188 & 801 \\
    Kuaishou & 7,176 & 10,728 & 12,530,806 & 31 \\
    \midrule
    Industrial & 170,000,000 & 310,000,000 & 118,000,000,000 & 27,452 \\
    \bottomrule
    \end{tabular}%
  }
  \vspace{-0.5em}
  \label{tab:stats}%
\end{table}%

\begin{table*}[tbp]
  \centering
  \caption{CTR prediction comparison results over \textit{five} trial runs ($\uparrow$: the higher, the better; $\downarrow$: the lower, the better). The best baseline(s) are highlighted with underlining.}
  \vspace{-0.5em}
  \resizebox{\linewidth}{!}{
    \begin{tabular}{c|ccc|ccc|ccc}
    \toprule
    \multirow{2}[4]{*}{Model} & \multicolumn{3}{c|}{MovieLens} & \multicolumn{3}{c|}{Electronics} & \multicolumn{3}{c}{Kuaishou} \\
\cmidrule{2-10}          & AUC ($\uparrow$)  & GAUC ($\uparrow$) & Logloss ($\downarrow$) & AUC ($\uparrow$) & GAUC ($\uparrow$) & Logloss ($\downarrow$) & AUC ($\uparrow$)  & GAUC ($\uparrow$) & Logloss ($\downarrow$) \\
    \midrule
    Wide\&Deep & 0.7237±0.0008 & 0.6922±0.0009 & 0.6072±0.0020 & 0.8242±0.0009 & 0.8247±0.0008 & 0.5132±0.0033 & 0.8202±0.0023 & 0.7761±0.0006 & 0.4922±0.0025 \\
    DeepFM & 0.7215±0.0015 & 0.6910±0.0011 & 0.6080±0.0026 & 0.8064±0.0028 & 0.8066±0.0028 & 0.5352±0.0081 & 0.8207±0.0014 & 0.7753±0.0007 & 0.4922±0.0023 \\
    AFM   & 0.7199±0.0008 & 0.6884±0.0007 & 0.6091±0.0013 & 0.7995±0.0008 & 0.7999±0.0009 & 0.5330±0.0008 & 0.8184±0.0034 & 0.7731±0.0049 & 0.4969±0.0041 \\
    NFM   & 0.7156±0.0039 & 0.6850±0.0042 & 0.6171±0.0078 & 0.8044±0.0009 & 0.8049±0.0009 & 0.5372±0.0033 & 0.8186±0.0045 & 0.7717±0.0022 & 0.4951±0.0040 \\
    \midrule
    DIN   & 0.7248±0.0010 & 0.6974±0.0005 & 0.6143±0.0043 & 0.8295±0.0026 & 0.8307±0.0030 & 0.5186±0.0028 & 0.8208±0.0019 & 0.7792±0.0005 & 0.4978±0.0031 \\
    DIEN  & 0.7262±0.0010 & 0.6958±0.0009 & 0.6112±0.0020 & 0.8313±0.0031 & 0.8323±0.0027 & 0.5167±0.0056 & \underline{0.8273}±0.0016 & 0.7783±0.0009 & 0.4943±0.0054 \\
    UBR4CTR & 0.7245±0.0002 & 0.6943±0.0010 & 0.6233±0.0076 & 0.8300±0.0005 & 0.8299±0.0006 & \underline{0.5056}±0.0007 & 0.8266±0.0005 & 0.7799±0.0006 & 0.4907±0.0020 \\
    SIM   & 0.7255±0.0014 & 0.6950±0.0012 & 0.6254±0.0094 & 0.8296±0.0033 & 0.8305±0.0031 & 0.5186±0.0062 & \underline{0.8273}±0.0005 & 0.7800±0.0005 & \underline{0.4906}±0.0021 \\
    \midrule
    PinSage & 0.7298±0.0017 & 0.7069±0.0017 & 0.6121±0.0039 & 0.8136±0.0027 & 0.8133±0.0027 & 0.5269±0.0078 & 0.8163±0.0019 & 0.7810±0.0006 & 0.5037±0.0041 \\
    LightGCN & 0.7305±0.0009 & 0.7077±0.0012 & 0.6122±0.0061 & \underline{0.8329}±0.0011 & \underline{0.8333}±0.0010 & 0.5101±0.0049 & 0.8139±0.0019 & 0.7803±0.0014 & 0.5068±0.0041 \\
    GLSM  & 0.7320±0.0003 & 0.7096±0.0007 & 0.6088±0.0035 & 0.8318±0.0026 & 0.8324±0.0026 & 0.5112±0.0066 & 0.8170±0.0012 & \underline{0.7811}±0.0004 & 0.5031±0.0059 \\
    GMT   & \underline{0.7353}±0.0014 & \underline{0.7097}±0.0010 & \underline{0.6003}±0.0023 & 0.8313±0.0020 & 0.8322±0.0024 & 0.5110±0.0083 & 0.8215±0.0018 & 0.7803±0.0017 & 0.4981±0.0020 \\
    \midrule
    \textbf{\model} & \textbf{0.7458}±0.0006 & \textbf{0.7198}±0.0007 & \textbf{0.5886}±0.0027 & \textbf{0.8444}±0.0009 & \textbf{0.8458}±0.0008 & \textbf{0.4892}±0.0040 & \textbf{0.8306}±0.0013 & \textbf{0.7813}±0.0010 & \textbf{0.4872}±0.0026 \\
    \bottomrule
    \end{tabular}%
    }
   \vspace{-0.3em}
  \label{tab:main_result}%
\end{table*}%

\subsection{Offline Evaluation (RQ1)}
In this subsection, we compare our proposed \model with twelve state-of-the-art baseline models on the four experimental datasets. The comparison results on the AUC and GAUC metrics are reported in Table \ref{tab:main_result} and Table \ref{tab:industrial_main}, with the following observations:

\textbf{\model can achieve significant improvements over state-of-the-art methods on all experimental datasets.} From the tables, we observe that the proposed \model achieves the highest AUC and GAUC performance and the lowest Logloss results.
Specifically, for the Logloss metric, \model outperforms the best baseline by 1.95\%, 4.10\%, 0.71\%, and 0.93\% on MovieLens, Electronics, Kuaishou, and the industrial dataset, respectively.
For all the AUC, GAUC, and Logloss metrics, \model brings effective gains of 1.00\%, 0.93\%, and 1.70\% on average respectively.
These comparison results verify that taking into account the graph information in a macro perspective of \model contributes to achieving better interest modeling and CTR prediction performance.

\textbf{The graph-based methods perform relatively well than other types of baseline models.}
Comparing the three main categories of baseline models, we can find the graph-based models (i.e. PinSage, LightGCN, GLSM, and GMT) obtain relatively better results than user interest modeling and feature interaction methods, which indicates that apart from the directly interacted neighborhood, incorporating high-order graph information can reflect the useful implicit preferences of the target user-item pair, and is significant for the overall CTR prediction performance.

\textbf{Increasing the modeling range through node sampling does not necessarily bring effective gains in all scenarios.}
These results show that applying node sampling-based methods (e.g. UBR4CTR, SIM, and GLSM) to consider behaviors does not consistently bring improvements to the performance. This suggests that modeling node interests by only searching and sampling similar nodes based on certain rules may not be accurate enough. Additionally, retrieving neighbors beyond the 1-hop using GLSM resulted in relatively better performance compared to UBR4CTR and SIM, also indicating that the higher-order interaction information is meaningful. The designed macro graph paradigm of \model avoids this issue, which is an important factor contributing to its optimal performance.

\begin{table}[tbp]
  \centering
  \caption{Comparison results on the industrial dataset.}
  \vspace{-0.9em}
  \resizebox{0.95\linewidth}{!}{
    \begin{tabular}{c|c|c|c}
    \toprule
    Industrial & AUC ($\uparrow$)  & GAUC ($\uparrow$)  & Logloss ($\downarrow$) \\
    \midrule
    Wide\&deep & 0.8123±0.0021 & 0.6908±0.0024 & 0.5223±0.0009 \\
    DeepFM & 0.8169±0.0012 & 0.6982±0.0036 & 0.5202±0.0018 \\
    AFM   & 0.8103±0.0008 & 0.6866±0.0021 & 0.5301±0.0020 \\
    NFM   & 0.8112±0.0031 & 0.6823±0.0043 & 0.5286±0.0032 \\
    \midrule
    DIN   & 0.8225±0.0017 & 0.6963±0.0013 & 0.5022±0.0012 \\
    DIEN  & 0.8231±0.0042 & 0.7008±0.0018 & 0.5009±0.0021 \\
    UBR4CTR & 0.8263±0.0037 & 0.7019±0.0032 & 0.4931±0.0019 \\
    SIM   & 0.8313±0.0025 & 0.7103±0.0010 & 0.4902±0.0008 \\
    \midrule
    PinSage  & 0.8289±0.0036 & 0.7086±0.0031 & 0.4917±0.0017 \\
    LightGCN & 0.8309±0.0006 & 0.7093±0.0018 & 0.4909±0.0012 \\
    GLSM  & 0.8326±0.0053 & 0.7149±0.0039 & 0.4887±0.0029 \\
    GMT   & \underline{0.8343}±0.0022 & \underline{0.7178}±0.0033 & \underline{0.4862}±0.0021 \\
    \midrule
    \textbf{\model} & \textbf{0.8408}±0.0019 & \textbf{0.7233}±0.0014 & \textbf{0.4817}±0.0013 \\
    \bottomrule
    \end{tabular}%
    }
    \vspace{-1.2em}
  \label{tab:industrial_main}%
\end{table}%

\subsection{Efficiency Study (RQ2)}
Since CTR prediction has to infer the user's intent in real-time and thus the computational efficiency of models is also an important evaluation factor~\cite{wang2019learning}. Hence, to verify the efficiency of \model, we compare the average online response time per user-item pair between \model and the well-performed and representative baselines: feature interaction-based model \textbf{Wide\&Deep}, user interest-based model \textbf{DIN} and the node searching scheme \textbf{SIM}, graph-based recursively convolution method \textbf{LightGCN} and graph transformer-based method \textbf{GMT}. Note that we present the online inference time on real-world recommender systems.

\begin{figure}[tbp]
    \centering
    \includegraphics[width=\linewidth, trim=0cm 0cm 0cm 0cm,clip]{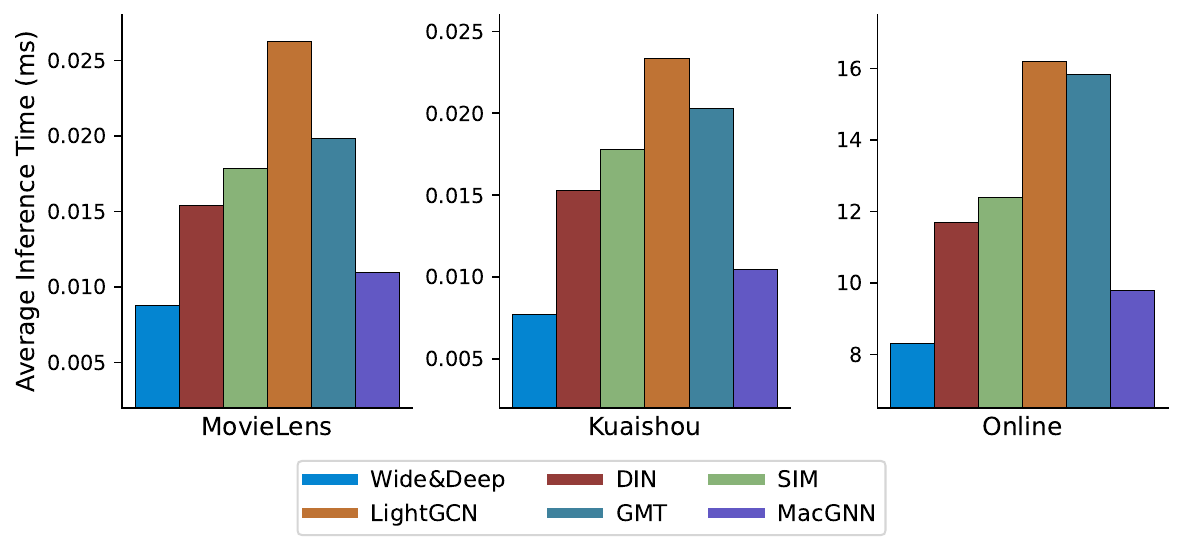}
    \vspace{-1em}
    \caption{Efficiency study of the model inference time.}
    \vspace{-0.5em}
    \label{fig:efficiency}
\end{figure}

The comparison result is shown in Figure \ref{fig:efficiency}. From the figure, we have the following observations: 
(i) The proposed model is almost as efficient as the fastest Wide\&Deep model. Apart from Wide\&Deep, our model achieves the best performance and efficiency among all user interest-based models and graph-based models.
(ii) While graph models employ sampling strategies to expedite the inference process, LightGCN and GMT are the two slowest models. Particularly on online platforms, LightGCN and GMT require nearly three times and two times the online inference time of MacGNN, leading to a significant online burden for billion-scale systems.

\begin{table}[tbp]
  \centering
  \caption{Ablation study results between \model with its four variants on MovieLens and Electronics.}
  \vspace{-0.8em}
  \resizebox{\linewidth}{!}{
    \begin{tabular}{cl|ccc}
    \toprule
    \multicolumn{2}{c|}{Variant} & AUC ($\uparrow$)   & GAUC ($\uparrow$)  & Logloss ($\downarrow$) \\
    \midrule
    \multirow{5}[2]{*}{\rotatebox{90}{MovieLens}} & \textbf{\model} & \textbf{0.7458}±0.0006 & \textbf{0.7198}±0.0007 & \textbf{0.5886}±0.0027 \\
    \cmidrule{2-5}      & \textit{w/o weighting} & 0.7396±0.0013 & 0.7132±0.0009 & 0.5923±0.0037 \\
          & \textit{w/o recent} & 0.7212±0.0005 & 0.6936±0.0009 & 0.6176±0.0052 \\
          \cmidrule{2-5} & \textit{w/o highorder} & 0.7401±0.0004 & 0.7126±0.0009 & 0.5929±0.0030 \\
          & \textit{w/o itemgraph} & 0.7239±0.0002 & 0.6871±0.0007 & 0.6073±0.0032 \\
    \midrule
    \multirow{5}[2]{*}{\rotatebox{90}{Electronics}} & \textbf{\model} & \textbf{0.8444}±0.0009 & \textbf{0.8458}±0.0008 & \textbf{0.4892}±0.0040 \\
    \cmidrule{2-5}      & \textit{w/o weighting} & 0.8418±0.0006 & 0.8417±0.0006 & 0.4938±0.0032 \\
          & \textit{w/o recent} & 0.8316±0.0010 & 0.8333±0.0008 & 0.5127±0.0037 \\
          \cmidrule{2-5} & \textit{w/o highorder} & 0.8302±0.0003 & 0.8319±0.0005 & 0.5083±0.0033 \\
          & \textit{w/o itemgraph} & 0.8189±0.0005 & 0.8199±0.0007 & 0.5259±0.0043 \\
    \bottomrule
    \end{tabular}%
    }
    \vspace{-1em}
  \label{tab:ablations}%
\end{table}%

\subsection{Ablation Study (RQ3)}
To verify the effectiveness of the key designed components and modeled information in \model, we conduct the ablation study by comparing \model with its four variants: 
(1) \textbf{\textit{w/o weighting}} removes the \textit{macro weight modeling} module in \model, which ignores the macro edge weights.
(2) \textbf{\textit{w/o recent}} removes the \textit{recent behavior modeling} scheme in \model, of which the short-term pattern modeling.
(3) \textbf{\textit{w/o highorder}} excludes the high-order graph information of the target user and item for \model training and the final prediction.
(4) \textbf{\textit{w/o itemgraph}} excludes the target item's graph information for \model training and prediction, which is largely ignored by previous works due to the efficiency trade-off.
From Table \ref{tab:ablations}., we have the following observations:

\textbf{Effectiveness of key designed components.}
(i) The lack of consideration of the macro edge weight results in the inferior performance of \textit{w/o weighting}, as the macro edge intensity can reflect the behavior pattern of users and items.
(ii) The removal of recent behavior may impact the recommendation performance of \textit{w/o recent} in comparison to \model. This underscores the importance of taking recent behaviors into account from a macro perspective.

\textbf{Effectiveness of key modeled information.}
(i) The decline in the performance of \textit{w/o highorder} relative to \model due to the neglect of high-order neighbors indicates the significance of graph information, and considering it from a macroscopic perspective is effective.
(ii) The substantial performance gap between \textit{w/o itemgraph} and \model highlights the significance of considering item-side graphs. Nonetheless, traditional CTR models tend to discard them due to computational constraints.

We have also conducted key parameter studies and case studies of \model, which are left in Appendix~\ref{sec:add_expers}.

\subsection{Online Evaluation (RQ4)}
We have deployed \model and conducted the online A/B test in Taobao’s homepage feed of Alibaba, one of the biggest online shopping platforms in China. 
The online performance is compared against the best-performed user interest-based model SIM and the sampling-based graph model GMT. 
The performance in Table~\ref{tab:abtest} is averaged over four consecutive weeks. 

\textit{Compared to SIM}, firstly, MacGNN demonstrates a performance improvement of $3.13\%$ for PCTR, $1.32\%$ for UCTR, and $5.13\%$ for GMV, suggesting that our model enhances users' willingness to engage with items and convert to purchases. 
Secondly, the Stay Time increases by $1.01\%$, indicating that MacGNN can effectively engage users, encouraging them to spend more time on the platform by catering to their comprehensive macro behavior interests.
Thirdly, MacGNN achieves a Response Time that is $20.97\%$ faster than SIM, showing that MacGNN achieves significantly improved performance and enhanced efficiency.

\textit{Compared to GMT}, MacGNN still demonstrates a performance improvement of $2.35\%$ for PCTR, $1.09\%$ for UCTR, and $3.53\%$ for GMV. This suggests that taking into account the complete macro behavior patterns of users and items can yield significantly better performance than considering only a small portion of sampled neighbors.
Furthermore, the Stay Time increases by $0.69\%$, indicating that MacGNN encourages users to stay by considering more comprehensive behavior patterns.
Lastly, MacGNN's Response Time is $38.13\%$ faster than SIM, confirming the efficiency of MAG.


\section{Related Work}

\subsection{Click-Through Rate Prediction}
Tradition models utilize feature interaction for CTR prediction~\cite{ijcai2021p636,zhu2021open}.
FM~\cite{rendle2010factorization} first introduces the latent vectors for 2-order feature interaction to address the feature sparsity.
Wide\&Deep~\cite{cheng2016wide} conducts feature interaction by a wide linear regression model and a deep feed-forward network with joint training.
DeepFM~\cite{guo2017deepfm} further replaces the linear regression in Wide\&Deep with FM to avoid feature engineering.
Recently, user interest-based models have achieved better CTR performance.
DIN~\cite{zhou2018deep} first designs a deep interest network with an attention mechanism between the user's behavior sequence and the target item.
DIEN~\cite{zhou2019deep} then enhances DIN with GRU~\cite{chung2014empirical} for user's evolution patterns mining.
Further, some studies aim to model longer behavior sequences with sampling~\cite{cao2022sampling,zhang2022clustering}.
UBR4CTR~\cite{qin2020user} samples the most relevant item from the user's long interaction behaviors.
Similarly, SIM~\cite{pi2020search} designs a two-stage paradigm, sampling relevant items and computing their attention score with the target, to reduce the scale of the user's complete behaviors. 


\begin{table}[tbp]
  \centering
  \caption{Results of online A/B tests in the industrial platform.}
  \resizebox{0.975\linewidth}{!}{
    \begin{tabular}{l|ccccc}
    \toprule
    A/B Test & PCTR & UCTR & GMV & StayTime & ResTime\\
    \midrule
    v.s. SIM & +3.13\% & +1.32\% & +5.13\% & +1.01\% & -20.97\%\\
    v.s. GMT & +2.35\% & +1.09\% & +3.53\% & +0.69\% & -38.13\%\\
    \bottomrule
    \end{tabular}%
    \vspace{-0.5em}
   }
  \label{tab:abtest}%
\end{table}%

\subsection{Graph Learning for Recommendation}
Recently, massive works have attempted to improve recommendation performance through graph learning methods~\cite{shengyuan2023differentiable,huang2023aldi,chen2021non}.
Typically, NGCF~\cite{wang2019neural} enhances traditional collaborative filtering with high-order graph information. 
LightGCN~\cite{he2020lightgcn} then removes the non-linear operation in NGCF, which is drawn from the observation of extensive experimental analysis.
These methods have been widely used for appropriate item recalling in industrial recommender systems. However, due to the strict requirements for time efficiency, they cannot be applied directly as CTR prediction models.
Then, some advances try to consider the graph information in the CTR scenario but they still maintain the node sampling paradigm.
GLSM~\cite{sun2022graph} conducts relevant node retrieval of the central user from the interaction graph for long-term interest modeling.
GMT~\cite{min2022neighbour} constructs a heterogeneous information network (HIN) with sampled various types of user interactions and designs a graph-masked transformer for user modeling.
NRCGI~\cite{bei2023nrcgi} further models the deeper graph information with cluster-scale sampled neighborhoods based on non-recursive aggregation.

\section{Conclusion}


In this paper, we introduce the \textit{Macro Recommendation Graph (MAG)} and \textit{Macro Graph Neural Networks (MacGNN)} for billion-scale recommender systems, offering a more suitable solution to the prevalent issues of computational complexity and sampling bias in conventional GNN recommendation models. By ingeniously grouping micro nodes into macro nodes, MAG allows for efficient computation, while MacGNN facilitates effective information aggregation and embedding refinement at a macro level. Demonstrating superior performance in both offline experiments and online A/B tests, and practically serving over a billion users in Alibaba, \model not only elevates the capability of predictive models in expansive digital environments but also paves the way for future research and optimizations in the realm of large-scale recommendation systems.

\begin{acks}
This work was supported in part by the Scientific Innovation 2030 Major Project for New Generation of AI (Grant No. 2020AAA0107300), Ministry of Science and Technology of the People’s Republic of China, and the National Natural Science Foundation of China (Grant No. 62272200, U22A2095, 61932010, 62172443).
\end{acks}

\bibliographystyle{ACM-Reference-Format}
\bibliography{reference.bib}


\begin{thebibliography}{43}


\ifx \showCODEN    \undefined \def \showCODEN     #1{\unskip}     \fi
\ifx \showDOI      \undefined \def \showDOI       #1{#1}\fi
\ifx \showISBNx    \undefined \def \showISBNx     #1{\unskip}     \fi
\ifx \showISBNxiii \undefined \def \showISBNxiii  #1{\unskip}     \fi
\ifx \showISSN     \undefined \def \showISSN      #1{\unskip}     \fi
\ifx \showLCCN     \undefined \def \showLCCN      #1{\unskip}     \fi
\ifx \shownote     \undefined \def \shownote      #1{#1}          \fi
\ifx \showarticletitle \undefined \def \showarticletitle #1{#1}   \fi
\ifx \showURL      \undefined \def \showURL       {\relax}        \fi
\providecommand\bibfield[2]{#2}
\providecommand\bibinfo[2]{#2}
\providecommand\natexlab[1]{#1}
\providecommand\showeprint[2][]{arXiv:#2}

\bibitem[Bei et~al\mbox{.}(2023a)]%
        {bei2023nrcgi}
\bibfield{author}{\bibinfo{person}{Yuanchen Bei}, \bibinfo{person}{Hao Chen}, \bibinfo{person}{Shengyuan Chen}, \bibinfo{person}{Xiao Huang}, \bibinfo{person}{Sheng Zhou}, {and} \bibinfo{person}{Feiran Huang}.} \bibinfo{year}{2023}\natexlab{a}.
\newblock \showarticletitle{Non-Recursive Cluster-Scale Graph Interacted Model for Click-Through Rate Prediction}. In \bibinfo{booktitle}{\emph{Proceedings of the 32nd ACM International Conference on Information and Knowledge Management}}. \bibinfo{pages}{3748--3752}.
\newblock


\bibitem[Bei et~al\mbox{.}(2023b)]%
        {bei2023reinforcement}
\bibfield{author}{\bibinfo{person}{Yuanchen Bei}, \bibinfo{person}{Sheng Zhou}, \bibinfo{person}{Qiaoyu Tan}, \bibinfo{person}{Hao Xu}, \bibinfo{person}{Hao Chen}, \bibinfo{person}{Zhao Li}, {and} \bibinfo{person}{Jiajun Bu}.} \bibinfo{year}{2023}\natexlab{b}.
\newblock \showarticletitle{Reinforcement Neighborhood Selection for Unsupervised Graph Anomaly Detection}. In \bibinfo{booktitle}{\emph{2023 IEEE International Conference on Data Mining (ICDM)}}.
\newblock


\bibitem[Cao et~al\mbox{.}(2022)]%
        {cao2022sampling}
\bibfield{author}{\bibinfo{person}{Yue Cao}, \bibinfo{person}{Xiaojiang Zhou}, \bibinfo{person}{Jiaqi Feng}, \bibinfo{person}{Peihao Huang}, \bibinfo{person}{Yao Xiao}, \bibinfo{person}{Dayao Chen}, {and} \bibinfo{person}{Sheng Chen}.} \bibinfo{year}{2022}\natexlab{}.
\newblock \showarticletitle{Sampling Is All You Need on Modeling Long-Term User Behaviors for CTR Prediction}. In \bibinfo{booktitle}{\emph{Proceedings of the 31st ACM International Conference on Information \& Knowledge Management}}. \bibinfo{pages}{2974--2983}.
\newblock


\bibitem[Chen et~al\mbox{.}(2021)]%
        {chen2021non}
\bibfield{author}{\bibinfo{person}{Hao Chen}, \bibinfo{person}{Zengde Deng}, \bibinfo{person}{Yue Xu}, {and} \bibinfo{person}{Zhoujun Li}.} \bibinfo{year}{2021}\natexlab{}.
\newblock \showarticletitle{Non-recursive graph convolutional networks}. In \bibinfo{booktitle}{\emph{ICASSP 2021-2021 IEEE International Conference on Acoustics, Speech and Signal Processing (ICASSP)}}. IEEE, \bibinfo{pages}{3205--3209}.
\newblock


\bibitem[Chen et~al\mbox{.}(2022)]%
        {chen2022gar}
\bibfield{author}{\bibinfo{person}{Hao Chen}, \bibinfo{person}{Zefan Wang}, \bibinfo{person}{Feiran Huang}, \bibinfo{person}{Xiao Huang}, \bibinfo{person}{Yue Xu}, \bibinfo{person}{Yishi Lin}, \bibinfo{person}{Peng He}, {and} \bibinfo{person}{Zhoujun Li}.} \bibinfo{year}{2022}\natexlab{}.
\newblock \showarticletitle{Generative adversarial framework for cold-start item recommendation}. In \bibinfo{booktitle}{\emph{Proceedings of the 45th International ACM SIGIR Conference on Research and Development in Information Retrieval}}. \bibinfo{pages}{2565--2571}.
\newblock


\bibitem[Chen et~al\mbox{.}(2016)]%
        {chen2016logloss}
\bibfield{author}{\bibinfo{person}{Junxuan Chen}, \bibinfo{person}{Baigui Sun}, \bibinfo{person}{Hao Li}, \bibinfo{person}{Hongtao Lu}, {and} \bibinfo{person}{Xian-Sheng Hua}.} \bibinfo{year}{2016}\natexlab{}.
\newblock \showarticletitle{Deep ctr prediction in display advertising}. In \bibinfo{booktitle}{\emph{Proceedings of the 24th ACM international conference on Multimedia}}. \bibinfo{pages}{811--820}.
\newblock


\bibitem[Chen et~al\mbox{.}(2023)]%
        {chen2023adap}
\bibfield{author}{\bibinfo{person}{Jiawei Chen}, \bibinfo{person}{Junkang Wu}, \bibinfo{person}{Jiancan Wu}, \bibinfo{person}{Xuezhi Cao}, \bibinfo{person}{Sheng Zhou}, {and} \bibinfo{person}{Xiangnan He}.} \bibinfo{year}{2023}\natexlab{}.
\newblock \showarticletitle{Adap-$\tau$: Adaptively Modulating Embedding Magnitude for Recommendation}. In \bibinfo{booktitle}{\emph{Proceedings of the ACM Web Conference 2023}}. \bibinfo{pages}{1085--1096}.
\newblock


\bibitem[Cheng et~al\mbox{.}(2016)]%
        {cheng2016wide}
\bibfield{author}{\bibinfo{person}{Heng-Tze Cheng}, \bibinfo{person}{Levent Koc}, \bibinfo{person}{Jeremiah Harmsen}, \bibinfo{person}{Tal Shaked}, \bibinfo{person}{Tushar Chandra}, \bibinfo{person}{Hrishi Aradhye}, \bibinfo{person}{Glen Anderson}, \bibinfo{person}{Greg Corrado}, \bibinfo{person}{Wei Chai}, \bibinfo{person}{Mustafa Ispir}, {et~al\mbox{.}}} \bibinfo{year}{2016}\natexlab{}.
\newblock \showarticletitle{Wide \& deep learning for recommender systems}. In \bibinfo{booktitle}{\emph{Proceedings of the 1st workshop on deep learning for recommender systems}}. \bibinfo{pages}{7--10}.
\newblock


\bibitem[Chung et~al\mbox{.}(2014)]%
        {chung2014empirical}
\bibfield{author}{\bibinfo{person}{Junyoung Chung}, \bibinfo{person}{Caglar Gulcehre}, \bibinfo{person}{KyungHyun Cho}, {and} \bibinfo{person}{Yoshua Bengio}.} \bibinfo{year}{2014}\natexlab{}.
\newblock \showarticletitle{Empirical evaluation of gated recurrent neural networks on sequence modeling}. In \bibinfo{booktitle}{\emph{NIPS 2014 Deep Learning Workshop}}.
\newblock


\bibitem[Covington et~al\mbox{.}(2016)]%
        {covington2016deep}
\bibfield{author}{\bibinfo{person}{Paul Covington}, \bibinfo{person}{Jay Adams}, {and} \bibinfo{person}{Emre Sargin}.} \bibinfo{year}{2016}\natexlab{}.
\newblock \showarticletitle{Deep neural networks for youtube recommendations}. In \bibinfo{booktitle}{\emph{Proceedings of the 10th ACM conference on recommender systems}}. \bibinfo{pages}{191--198}.
\newblock


\bibitem[Deng et~al\mbox{.}(2018)]%
        {deng2018ad}
\bibfield{author}{\bibinfo{person}{Weiwei Deng}, \bibinfo{person}{Xiaoliang Ling}, \bibinfo{person}{Yang Qi}, \bibinfo{person}{Tunzi Tan}, \bibinfo{person}{Eren Manavoglu}, {and} \bibinfo{person}{Qi Zhang}.} \bibinfo{year}{2018}\natexlab{}.
\newblock \showarticletitle{Ad click prediction in sequence with long short-term memory networks: an externality-aware model}. In \bibinfo{booktitle}{\emph{The 41st International ACM SIGIR Conference on Research \& Development in Information Retrieval}}. \bibinfo{pages}{1065--1068}.
\newblock


\bibitem[Fawcett(2006)]%
        {fawcett2006introduction}
\bibfield{author}{\bibinfo{person}{Tom Fawcett}.} \bibinfo{year}{2006}\natexlab{}.
\newblock \showarticletitle{An introduction to ROC analysis}.
\newblock \bibinfo{journal}{\emph{Pattern recognition letters}} \bibinfo{volume}{27}, \bibinfo{number}{8} (\bibinfo{year}{2006}), \bibinfo{pages}{861--874}.
\newblock


\bibitem[Gao et~al\mbox{.}(2022)]%
        {gao2022kuairec}
\bibfield{author}{\bibinfo{person}{Chongming Gao}, \bibinfo{person}{Shijun Li}, \bibinfo{person}{Wenqiang Lei}, \bibinfo{person}{Jiawei Chen}, \bibinfo{person}{Biao Li}, \bibinfo{person}{Peng Jiang}, \bibinfo{person}{Xiangnan He}, \bibinfo{person}{Jiaxin Mao}, {and} \bibinfo{person}{Tat-Seng Chua}.} \bibinfo{year}{2022}\natexlab{}.
\newblock \showarticletitle{KuaiRec: A Fully-observed Dataset and Insights for Evaluating Recommender Systems}. In \bibinfo{booktitle}{\emph{Proceedings of the 31st ACM International Conference on Information \& Knowledge Management}}. \bibinfo{pages}{540--550}.
\newblock


\bibitem[Glorot and Bengio(2010)]%
        {glorot2010understanding}
\bibfield{author}{\bibinfo{person}{Xavier Glorot} {and} \bibinfo{person}{Yoshua Bengio}.} \bibinfo{year}{2010}\natexlab{}.
\newblock \showarticletitle{Understanding the difficulty of training deep feedforward neural networks}. In \bibinfo{booktitle}{\emph{Proceedings of the thirteenth international conference on artificial intelligence and statistics}}. JMLR Workshop and Conference Proceedings, \bibinfo{pages}{249--256}.
\newblock


\bibitem[Guo et~al\mbox{.}(2017)]%
        {guo2017deepfm}
\bibfield{author}{\bibinfo{person}{Huifeng Guo}, \bibinfo{person}{Ruiming Tang}, \bibinfo{person}{Yunming Ye}, \bibinfo{person}{Zhenguo Li}, {and} \bibinfo{person}{Xiuqiang He}.} \bibinfo{year}{2017}\natexlab{}.
\newblock \showarticletitle{DeepFM: a factorization-machine based neural network for CTR prediction}. In \bibinfo{booktitle}{\emph{Proceedings of the 26th International Joint Conference on Artificial Intelligence}}. \bibinfo{pages}{1725--1731}.
\newblock


\bibitem[Hamerly and Elkan(2003)]%
        {hamerly2003learning}
\bibfield{author}{\bibinfo{person}{Greg Hamerly} {and} \bibinfo{person}{Charles Elkan}.} \bibinfo{year}{2003}\natexlab{}.
\newblock \showarticletitle{Learning the k in k-means}.
\newblock \bibinfo{journal}{\emph{Advances in neural information processing systems}}  \bibinfo{volume}{16} (\bibinfo{year}{2003}).
\newblock


\bibitem[Harper and Konstan(2015)]%
        {harper2015movielens}
\bibfield{author}{\bibinfo{person}{F~Maxwell Harper} {and} \bibinfo{person}{Joseph~A Konstan}.} \bibinfo{year}{2015}\natexlab{}.
\newblock \showarticletitle{The movielens datasets: History and context}.
\newblock \bibinfo{journal}{\emph{Acm transactions on interactive intelligent systems (tiis)}} \bibinfo{volume}{5}, \bibinfo{number}{4} (\bibinfo{year}{2015}), \bibinfo{pages}{1--19}.
\newblock


\bibitem[He and Chua(2017)]%
        {he2017nfm}
\bibfield{author}{\bibinfo{person}{Xiangnan He} {and} \bibinfo{person}{Tat-Seng Chua}.} \bibinfo{year}{2017}\natexlab{}.
\newblock \showarticletitle{Neural factorization machines for sparse predictive analytics}. In \bibinfo{booktitle}{\emph{Proceedings of the 40th International ACM SIGIR conference on Research and Development in Information Retrieval}}. \bibinfo{pages}{355--364}.
\newblock


\bibitem[He et~al\mbox{.}(2020)]%
        {he2020lightgcn}
\bibfield{author}{\bibinfo{person}{Xiangnan He}, \bibinfo{person}{Kuan Deng}, \bibinfo{person}{Xiang Wang}, \bibinfo{person}{Yan Li}, \bibinfo{person}{Yongdong Zhang}, {and} \bibinfo{person}{Meng Wang}.} \bibinfo{year}{2020}\natexlab{}.
\newblock \showarticletitle{Lightgcn: Simplifying and powering graph convolution network for recommendation}. In \bibinfo{booktitle}{\emph{Proceedings of the 43rd International ACM SIGIR conference on research and development in Information Retrieval}}. \bibinfo{pages}{639--648}.
\newblock


\bibitem[Huang et~al\mbox{.}(2023)]%
        {huang2023aldi}
\bibfield{author}{\bibinfo{person}{Feiran Huang}, \bibinfo{person}{Zefan Wang}, \bibinfo{person}{Xiao Huang}, \bibinfo{person}{Yufeng Qian}, \bibinfo{person}{Zhetao Li}, {and} \bibinfo{person}{Hao Chen}.} \bibinfo{year}{2023}\natexlab{}.
\newblock \showarticletitle{Aligning Distillation For Cold-Start Item Recommendation}. In \bibinfo{booktitle}{\emph{Proceedings of the 46th International ACM SIGIR Conference on Research and Development in Information Retrieval}}. \bibinfo{pages}{1147–1157}.
\newblock


\bibitem[Kingma and Ba(2015)]%
        {kingma2014adam}
\bibfield{author}{\bibinfo{person}{Diederik~P Kingma} {and} \bibinfo{person}{Jimmy Ba}.} \bibinfo{year}{2015}\natexlab{}.
\newblock \showarticletitle{Adam: A method for stochastic optimization}. In \bibinfo{booktitle}{\emph{International Conference on Learning Representations}}.
\newblock


\bibitem[McAuley et~al\mbox{.}(2015)]%
        {mcauley2015image}
\bibfield{author}{\bibinfo{person}{Julian McAuley}, \bibinfo{person}{Christopher Targett}, \bibinfo{person}{Qinfeng Shi}, {and} \bibinfo{person}{Anton Van Den~Hengel}.} \bibinfo{year}{2015}\natexlab{}.
\newblock \showarticletitle{Image-based recommendations on styles and substitutes}. In \bibinfo{booktitle}{\emph{Proceedings of the 38th international ACM SIGIR conference on research and development in information retrieval}}. \bibinfo{pages}{43--52}.
\newblock


\bibitem[Min et~al\mbox{.}(2022)]%
        {min2022neighbour}
\bibfield{author}{\bibinfo{person}{Erxue Min}, \bibinfo{person}{Yu Rong}, \bibinfo{person}{Tingyang Xu}, \bibinfo{person}{Yatao Bian}, \bibinfo{person}{Da Luo}, \bibinfo{person}{Kangyi Lin}, \bibinfo{person}{Junzhou Huang}, \bibinfo{person}{Sophia Ananiadou}, {and} \bibinfo{person}{Peilin Zhao}.} \bibinfo{year}{2022}\natexlab{}.
\newblock \showarticletitle{Neighbour interaction based click-through rate prediction via graph-masked transformer}. In \bibinfo{booktitle}{\emph{Proceedings of the 45th International ACM SIGIR Conference on Research and Development in Information Retrieval}}. \bibinfo{pages}{353--362}.
\newblock


\bibitem[Pfadler et~al\mbox{.}(2020)]%
        {pfadler2020billion}
\bibfield{author}{\bibinfo{person}{Andreas Pfadler}, \bibinfo{person}{Huan Zhao}, \bibinfo{person}{Jizhe Wang}, \bibinfo{person}{Lifeng Wang}, \bibinfo{person}{Pipei Huang}, {and} \bibinfo{person}{Dik~Lun Lee}.} \bibinfo{year}{2020}\natexlab{}.
\newblock \showarticletitle{Billion-scale recommendation with heterogeneous side information at taobao}. In \bibinfo{booktitle}{\emph{2020 IEEE 36th International Conference on Data Engineering (ICDE)}}. IEEE, \bibinfo{pages}{1667--1676}.
\newblock


\bibitem[Pi et~al\mbox{.}(2020)]%
        {pi2020search}
\bibfield{author}{\bibinfo{person}{Qi Pi}, \bibinfo{person}{Guorui Zhou}, \bibinfo{person}{Yujing Zhang}, \bibinfo{person}{Zhe Wang}, \bibinfo{person}{Lejian Ren}, \bibinfo{person}{Ying Fan}, \bibinfo{person}{Xiaoqiang Zhu}, {and} \bibinfo{person}{Kun Gai}.} \bibinfo{year}{2020}\natexlab{}.
\newblock \showarticletitle{Search-based user interest modeling with lifelong sequential behavior data for click-through rate prediction}. In \bibinfo{booktitle}{\emph{Proceedings of the 29th ACM International Conference on Information \& Knowledge Management}}. \bibinfo{pages}{2685--2692}.
\newblock


\bibitem[Qin et~al\mbox{.}(2020)]%
        {qin2020user}
\bibfield{author}{\bibinfo{person}{Jiarui Qin}, \bibinfo{person}{Weinan Zhang}, \bibinfo{person}{Xin Wu}, \bibinfo{person}{Jiarui Jin}, \bibinfo{person}{Yuchen Fang}, {and} \bibinfo{person}{Yong Yu}.} \bibinfo{year}{2020}\natexlab{}.
\newblock \showarticletitle{User behavior retrieval for click-through rate prediction}. In \bibinfo{booktitle}{\emph{Proceedings of the 43rd International ACM SIGIR Conference on Research and Development in Information Retrieval}}. \bibinfo{pages}{2347--2356}.
\newblock


\bibitem[Rendle(2010)]%
        {rendle2010factorization}
\bibfield{author}{\bibinfo{person}{Steffen Rendle}.} \bibinfo{year}{2010}\natexlab{}.
\newblock \showarticletitle{Factorization machines}. In \bibinfo{booktitle}{\emph{2010 IEEE International conference on data mining}}. IEEE, \bibinfo{pages}{995--1000}.
\newblock


\bibitem[Shengyuan et~al\mbox{.}(2023)]%
        {shengyuan2023differentiable}
\bibfield{author}{\bibinfo{person}{Chen Shengyuan}, \bibinfo{person}{Yunfeng Cai}, \bibinfo{person}{Huang Fang}, \bibinfo{person}{Xiao Huang}, {and} \bibinfo{person}{Mingming Sun}.} \bibinfo{year}{2023}\natexlab{}.
\newblock \showarticletitle{Differentiable Neuro-Symbolic Reasoning on Large-Scale Knowledge Graphs}. In \bibinfo{booktitle}{\emph{Thirty-seventh Conference on Neural Information Processing Systems}}.
\newblock


\bibitem[Sun et~al\mbox{.}(2022)]%
        {sun2022graph}
\bibfield{author}{\bibinfo{person}{Huinan Sun}, \bibinfo{person}{Guangliang Yu}, \bibinfo{person}{Pengye Zhang}, \bibinfo{person}{Bo Zhang}, \bibinfo{person}{Xingxing Wang}, {and} \bibinfo{person}{Dong Wang}.} \bibinfo{year}{2022}\natexlab{}.
\newblock \showarticletitle{Graph Based Long-Term And Short-Term Interest Model for Click-Through Rate Prediction}. In \bibinfo{booktitle}{\emph{Proceedings of the 31st ACM International Conference on Information \& Knowledge Management}}. \bibinfo{pages}{1818--1826}.
\newblock


\bibitem[Wang et~al\mbox{.}(2019b)]%
        {wang2019learning}
\bibfield{author}{\bibinfo{person}{Weixun Wang}, \bibinfo{person}{Junqi Jin}, \bibinfo{person}{Jianye Hao}, \bibinfo{person}{Chunjie Chen}, \bibinfo{person}{Chuan Yu}, \bibinfo{person}{Weinan Zhang}, \bibinfo{person}{Jun Wang}, \bibinfo{person}{Xiaotian Hao}, \bibinfo{person}{Yixi Wang}, \bibinfo{person}{Han Li}, {et~al\mbox{.}}} \bibinfo{year}{2019}\natexlab{b}.
\newblock \showarticletitle{Learning adaptive display exposure for real-time advertising}. In \bibinfo{booktitle}{\emph{Proceedings of the 28th ACM International Conference on Information and Knowledge Management}}. \bibinfo{pages}{2595--2603}.
\newblock


\bibitem[Wang et~al\mbox{.}(2019a)]%
        {wang2019neural}
\bibfield{author}{\bibinfo{person}{Xiang Wang}, \bibinfo{person}{Xiangnan He}, \bibinfo{person}{Meng Wang}, \bibinfo{person}{Fuli Feng}, {and} \bibinfo{person}{Tat-Seng Chua}.} \bibinfo{year}{2019}\natexlab{a}.
\newblock \showarticletitle{Neural graph collaborative filtering}. In \bibinfo{booktitle}{\emph{Proceedings of the 42nd international ACM SIGIR conference on Research and development in Information Retrieval}}. \bibinfo{pages}{165--174}.
\newblock


\bibitem[Wu et~al\mbox{.}(2022)]%
        {wu2022graph}
\bibfield{author}{\bibinfo{person}{Shiwen Wu}, \bibinfo{person}{Fei Sun}, \bibinfo{person}{Wentao Zhang}, \bibinfo{person}{Xu Xie}, {and} \bibinfo{person}{Bin Cui}.} \bibinfo{year}{2022}\natexlab{}.
\newblock \showarticletitle{Graph neural networks in recommender systems: a survey}.
\newblock \bibinfo{journal}{\emph{Comput. Surveys}} \bibinfo{volume}{55}, \bibinfo{number}{5} (\bibinfo{year}{2022}), \bibinfo{pages}{1--37}.
\newblock


\bibitem[Xiao et~al\mbox{.}(2017)]%
        {xiao2017attentional}
\bibfield{author}{\bibinfo{person}{Jun Xiao}, \bibinfo{person}{Hao Ye}, \bibinfo{person}{Xiangnan He}, \bibinfo{person}{Hanwang Zhang}, \bibinfo{person}{Fei Wu}, {and} \bibinfo{person}{Tat-Seng Chua}.} \bibinfo{year}{2017}\natexlab{}.
\newblock \showarticletitle{Attentional factorization machines: learning the weight of feature interactions via attention networks}. In \bibinfo{booktitle}{\emph{Proceedings of the 26th International Joint Conference on Artificial Intelligence}}. \bibinfo{pages}{3119--3125}.
\newblock


\bibitem[Xu et~al\mbox{.}(2023)]%
        {xu2023multi}
\bibfield{author}{\bibinfo{person}{Yue Xu}, \bibinfo{person}{Hao Chen}, \bibinfo{person}{Zefan Wang}, \bibinfo{person}{Jianwen Yin}, \bibinfo{person}{Qijie Shen}, \bibinfo{person}{Dimin Wang}, \bibinfo{person}{Feiran Huang}, \bibinfo{person}{Lixiang Lai}, \bibinfo{person}{Tao Zhuang}, \bibinfo{person}{Junfeng Ge}, {et~al\mbox{.}}} \bibinfo{year}{2023}\natexlab{}.
\newblock \showarticletitle{Multi-factor Sequential Re-ranking with Perception-Aware Diversification}. In \bibinfo{booktitle}{\emph{Proceedings of the 29th ACM SIGKDD Conference on Knowledge Discovery and Data Mining}}. \bibinfo{pages}{5327–5337}.
\newblock


\bibitem[Ying et~al\mbox{.}(2018)]%
        {ying2018graph}
\bibfield{author}{\bibinfo{person}{Rex Ying}, \bibinfo{person}{Ruining He}, \bibinfo{person}{Kaifeng Chen}, \bibinfo{person}{Pong Eksombatchai}, \bibinfo{person}{William~L Hamilton}, {and} \bibinfo{person}{Jure Leskovec}.} \bibinfo{year}{2018}\natexlab{}.
\newblock \showarticletitle{Graph convolutional neural networks for web-scale recommender systems}. In \bibinfo{booktitle}{\emph{Proceedings of the 24th ACM SIGKDD international conference on knowledge discovery \& data mining}}. \bibinfo{pages}{974--983}.
\newblock


\bibitem[Zangerle and Bauer(2022)]%
        {zangerle2022evaluating}
\bibfield{author}{\bibinfo{person}{Eva Zangerle} {and} \bibinfo{person}{Christine Bauer}.} \bibinfo{year}{2022}\natexlab{}.
\newblock \showarticletitle{Evaluating recommender systems: survey and framework}.
\newblock \bibinfo{journal}{\emph{Comput. Surveys}} \bibinfo{volume}{55}, \bibinfo{number}{8} (\bibinfo{year}{2022}), \bibinfo{pages}{1--38}.
\newblock


\bibitem[Zhang et~al\mbox{.}(2021)]%
        {ijcai2021p636}
\bibfield{author}{\bibinfo{person}{Weinan Zhang}, \bibinfo{person}{Jiarui Qin}, \bibinfo{person}{Wei Guo}, \bibinfo{person}{Ruiming Tang}, {and} \bibinfo{person}{Xiuqiang He}.} \bibinfo{year}{2021}\natexlab{}.
\newblock \showarticletitle{Deep Learning for Click-Through Rate Estimation}. In \bibinfo{booktitle}{\emph{Proceedings of the Thirtieth International Joint Conference on Artificial Intelligence, {IJCAI-21}}}. \bibinfo{pages}{4695--4703}.
\newblock


\bibitem[Zhang et~al\mbox{.}(2022)]%
        {zhang2022clustering}
\bibfield{author}{\bibinfo{person}{Yuren Zhang}, \bibinfo{person}{Enhong Chen}, \bibinfo{person}{Binbin Jin}, \bibinfo{person}{Hao Wang}, \bibinfo{person}{Min Hou}, \bibinfo{person}{Wei Huang}, {and} \bibinfo{person}{Runlong Yu}.} \bibinfo{year}{2022}\natexlab{}.
\newblock \showarticletitle{Clustering based behavior sampling with long sequential data for CTR prediction}. In \bibinfo{booktitle}{\emph{Proceedings of the 45th International ACM SIGIR Conference on Research and Development in Information Retrieval}}. \bibinfo{pages}{2195--2200}.
\newblock


\bibitem[Zhou et~al\mbox{.}(2021)]%
        {zhou2021contrastive}
\bibfield{author}{\bibinfo{person}{Chang Zhou}, \bibinfo{person}{Jianxin Ma}, \bibinfo{person}{Jianwei Zhang}, \bibinfo{person}{Jingren Zhou}, {and} \bibinfo{person}{Hongxia Yang}.} \bibinfo{year}{2021}\natexlab{}.
\newblock \showarticletitle{Contrastive learning for debiased candidate generation in large-scale recommender systems}. In \bibinfo{booktitle}{\emph{Proceedings of the 27th ACM SIGKDD Conference on Knowledge Discovery \& Data Mining}}. \bibinfo{pages}{3985--3995}.
\newblock


\bibitem[Zhou et~al\mbox{.}(2019)]%
        {zhou2019deep}
\bibfield{author}{\bibinfo{person}{Guorui Zhou}, \bibinfo{person}{Na Mou}, \bibinfo{person}{Ying Fan}, \bibinfo{person}{Qi Pi}, \bibinfo{person}{Weijie Bian}, \bibinfo{person}{Chang Zhou}, \bibinfo{person}{Xiaoqiang Zhu}, {and} \bibinfo{person}{Kun Gai}.} \bibinfo{year}{2019}\natexlab{}.
\newblock \showarticletitle{Deep interest evolution network for click-through rate prediction}. In \bibinfo{booktitle}{\emph{Proceedings of the AAAI conference on artificial intelligence}}, Vol.~\bibinfo{volume}{33}. \bibinfo{pages}{5941--5948}.
\newblock


\bibitem[Zhou et~al\mbox{.}(2018)]%
        {zhou2018deep}
\bibfield{author}{\bibinfo{person}{Guorui Zhou}, \bibinfo{person}{Xiaoqiang Zhu}, \bibinfo{person}{Chenru Song}, \bibinfo{person}{Ying Fan}, \bibinfo{person}{Han Zhu}, \bibinfo{person}{Xiao Ma}, \bibinfo{person}{Yanghui Yan}, \bibinfo{person}{Junqi Jin}, \bibinfo{person}{Han Li}, {and} \bibinfo{person}{Kun Gai}.} \bibinfo{year}{2018}\natexlab{}.
\newblock \showarticletitle{Deep interest network for click-through rate prediction}. In \bibinfo{booktitle}{\emph{Proceedings of the 24th ACM SIGKDD International Conference on Knowledge Discovery \& Data Mining}}. \bibinfo{pages}{1059--1068}.
\newblock


\bibitem[Zhou et~al\mbox{.}(2010)]%
        {zhou2010impact}
\bibfield{author}{\bibinfo{person}{Renjie Zhou}, \bibinfo{person}{Samamon Khemmarat}, {and} \bibinfo{person}{Lixin Gao}.} \bibinfo{year}{2010}\natexlab{}.
\newblock \showarticletitle{The impact of YouTube recommendation system on video views}. In \bibinfo{booktitle}{\emph{Proceedings of the 10th ACM SIGCOMM conference on Internet measurement}}. \bibinfo{pages}{404--410}.
\newblock


\bibitem[Zhu et~al\mbox{.}(2021)]%
        {zhu2021open}
\bibfield{author}{\bibinfo{person}{Jieming Zhu}, \bibinfo{person}{Jinyang Liu}, \bibinfo{person}{Shuai Yang}, \bibinfo{person}{Qi Zhang}, {and} \bibinfo{person}{Xiuqiang He}.} \bibinfo{year}{2021}\natexlab{}.
\newblock \showarticletitle{Open benchmarking for click-through rate prediction}. In \bibinfo{booktitle}{\emph{Proceedings of the 30th ACM International Conference on Information \& Knowledge Management}}. \bibinfo{pages}{2759--2769}.
\newblock


\end{thebibliography}

\appendix
\section{Dataset Details}\label{sec:app_data}
We adopt both three publicly available datasets and a billion-scale industrial dataset for offline evaluation.
The detailed description and preprocessing manner of the datasets are as follows:

\textbf{MovieLens Dataset}\footnote{https://grouplens.org/datasets/movielens/10m}~\cite{harper2015movielens} contains 71,567 users, 10,681 movies, and 10,000,054 interactions of users' ratings to the movies. To make the rating interactions suitable for the CTR prediction task, we follow the previous works~\cite{zhou2018deep} to transform the rating interactions into clicked and non-clicked relationships, which label the samples with rating values that greater than or equal to 4 to be positive and the rest to be negative.

\textbf{Electronics Dataset}\footnote{https://jmcauley.ucsd.edu/data/amazon}~\cite{mcauley2015image} is a subset of Amazon Dataset, which contains product reviews and metadata from Amazon. It contains 192,403 users, 63,001 items, and 1,689,188 interactions. We treat all the user reviews as user click behaviors, which is widely used in the related works~\cite{zhou2018deep,zhou2019deep}.


\textbf{Kuaishou Dataset}\footnote{https://kuairec.com}~\cite{gao2022kuairec} is a real-world dataset collected from the recommendation logs of the video-sharing mobile app Kuaishou. It contains 7,176 users, 10,728 videos, and 12,530,806 interactions. We regard the samples with video play time account for more than 50\% of the total time to be truly clicked videos.

\textbf{Industrial Dataset} is a billion-scale dataset collected from Alibaba Taobao's homepage, one of the largest e-commerce recommendation applications, involving billions scale of users and items. The industrial dataset contains both positive and negative interactions (e.g., impression without user clicks) such that negative sampling is not needed. There are over 118 billion instances and each user has around 938 recent behaviors on average, which is much longer than the sequences from the public dataset. Following SIM~\cite{pi2020search}, we use the instances of the past two weeks as the training set and the instances of the next day as the test set. 

For public datasets, the number of macro user clusters is set as 20 and the macro item cluster utilizes item categories for simplicity. For the industrial dataset, the number of macro user clusters is set as 200 and the number of macro item clusters is set as 300.

\section{Baseline Details}\label{sec:app_baseline}
We compare our proposed \model with twelve representative state-of-the-art CTR prediction models as follows.

\textit{\textbf{Feature Interaction-based Methods:}}
(i) \textbf{Wide\&Deep}~\cite{cheng2016wide} is widely used in real industrial applications. It consists of a wide module and a deep module to discover and extract the correlation and nonlinear relations between features.
(ii) \textbf{DeepFM}~\cite{guo2017deepfm} is a variant model of Wide\&Deep, which imposes a factorization machine (FM)~\cite{rendle2010factorization} as a wide part avoiding manufactured feature engineering.
(iii) \textbf{AFM}~\cite{xiao2017attentional} improves feature interactions by discriminating the different importance via an attention network. 
(iv) \textbf{NFM}~\cite{he2017nfm} introduces the bi-interaction pooling to deepen FM for learning higher-order and non-linear feature interactions.

\textit{\textbf{User Interest Modeling-based Methods:}}
(i) \textbf{DIN}~\cite{zhou2018deep} is the first model that uses an attention mechanism to extract user interest representation from truncated historical user behaviors in CTR prediction.
(ii) \textbf{DIEN}~\cite{zhou2019deep} is an improved version of DIN, which uses a two-layer RNNs module enhanced with the attention mechanism to capture the evolving user interests. 
(iii) \textbf{UBR4CTR}~\cite{qin2020user} proposes a search engine-based method to retrieve more relevant and appropriate behavioral data in long user sequential behaviors for model training.
(iv) \textbf{SIM}~\cite{pi2020search} uses two cascaded search units to extract user interests, which has a better ability to model long sequential behavior data in both scalability and performance in the CTR prediction.

\textit{\textbf{Graph-based Methods:}}
(i) \textbf{PinSage}~\cite{ying2018graph} is a representative graph-based web-scale recommendation model, which conducts inductive graph aggregation on the sampled user/item nodes. We concatenate and feed the trained embeddings by PinSage into the widely employed prediction layer to fit the CTR prediction scenario.
(ii) \textbf{LightGCN}~\cite{he2020lightgcn} is a simplified collaborative filtering model design by including only the most essential components in GCN for recommendation. Since it is a collaborative filtering model, the trained embeddings are also fed into the prediction layer as PinSage for the CTR prediction.
(iii) \textbf{GLSM}~\cite{sun2022graph} is a sampling-based model to introduce graph information, which consists of a multi-interest graph structure for capturing the long-term patterns and a sequence model for modeling the short-term information.
(iv) \textbf{GMT}~\cite{min2022neighbour} is also a sampling-based state-of-the-art graph model for CTR prediction with a graph-masked transformer to learn different kinds of interactions on the heterogeneous information network among the constructed neighborhood nodes.

\begin{figure}[tbp]
    \centering
    \includegraphics[width=0.985\linewidth, trim=0cm 0cm 0cm 0cm,clip]{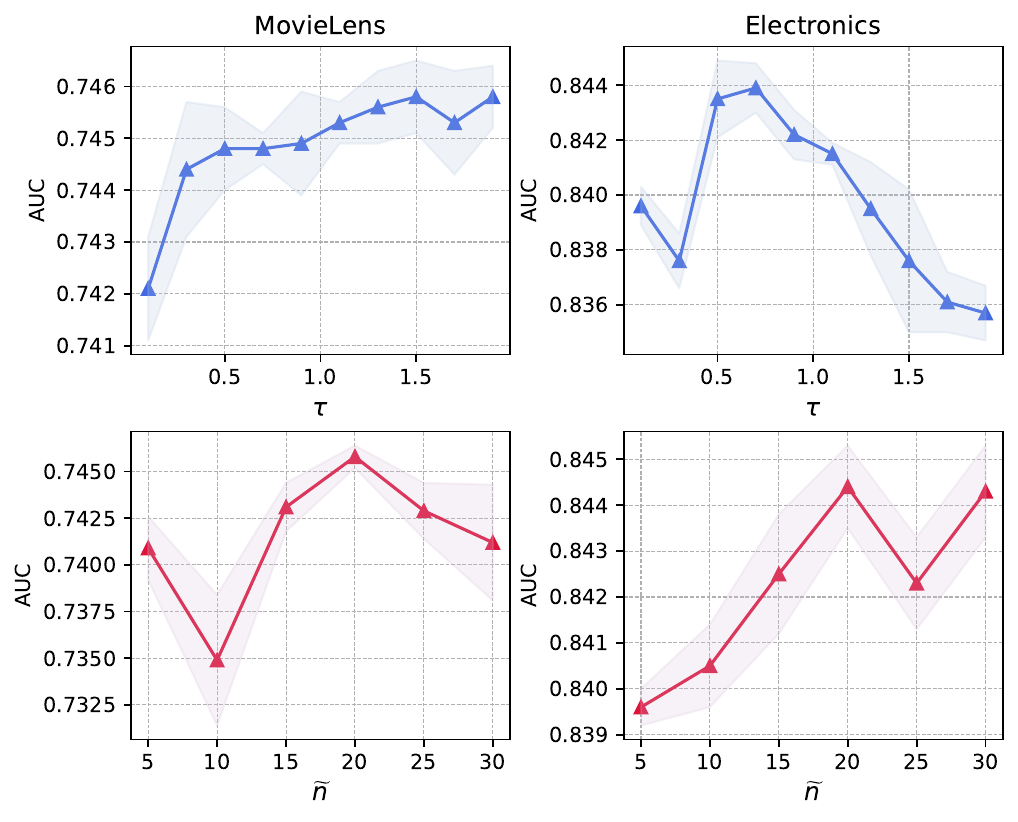}
    \caption{Parameter study of temperature parameter $\tau$ and macro node number $\widetilde{n}$ on MovieLens and Electronics.}
    \label{fig:param_tau_k}
\end{figure}

\section{Additional Experiments}\label{sec:add_expers}
\subsection{Parameter Analysis}
\subsubsection{\textbf{Effect of Temperature Parameter}}
We investigate the effect of the temperature parameter $\tau$ in macro node weighting with the range of 0.1 to 1.9 with a step size of 0.2 as illustrated in Figure \ref{fig:param_tau_k}.
We can observe from the results that a too-small weighting value of $\tau$ will cause poor performance. Furthermore, the suitable value of $\tau$ for MovieLens is larger than 1 while for Electronics is smaller than 1, one possible reason is that the temperature of \model should be set smaller on more sparse datasets.

\subsubsection{\textbf{Effect of Macro Node Number}}
We also evaluate the impact of different macro user numbers $\widetilde{n}$ under the behavior pattern grouping and fixed utilize of category as item grouping to avoid the impact of multiple variables.
From the second row of Figure \ref{fig:param_tau_k}, we can find that the too-small cluster number will lead to too coarsen user segmentation and result in poor results.
In addition, choosing a relatively appropriate number of clusters, such as 20, can bring good enough performance of \model on the public datasets and this macro node number is much smaller than the micro interaction scale, and also much smaller than the sequence length of previous user interest modeling works~\cite{zhou2018deep,zhou2019deep}.

\subsection{Case Study}
\begin{figure}[tbp]
    \centering
    \includegraphics[width= \linewidth, trim=0cm 0cm 0cm 0cm,clip]{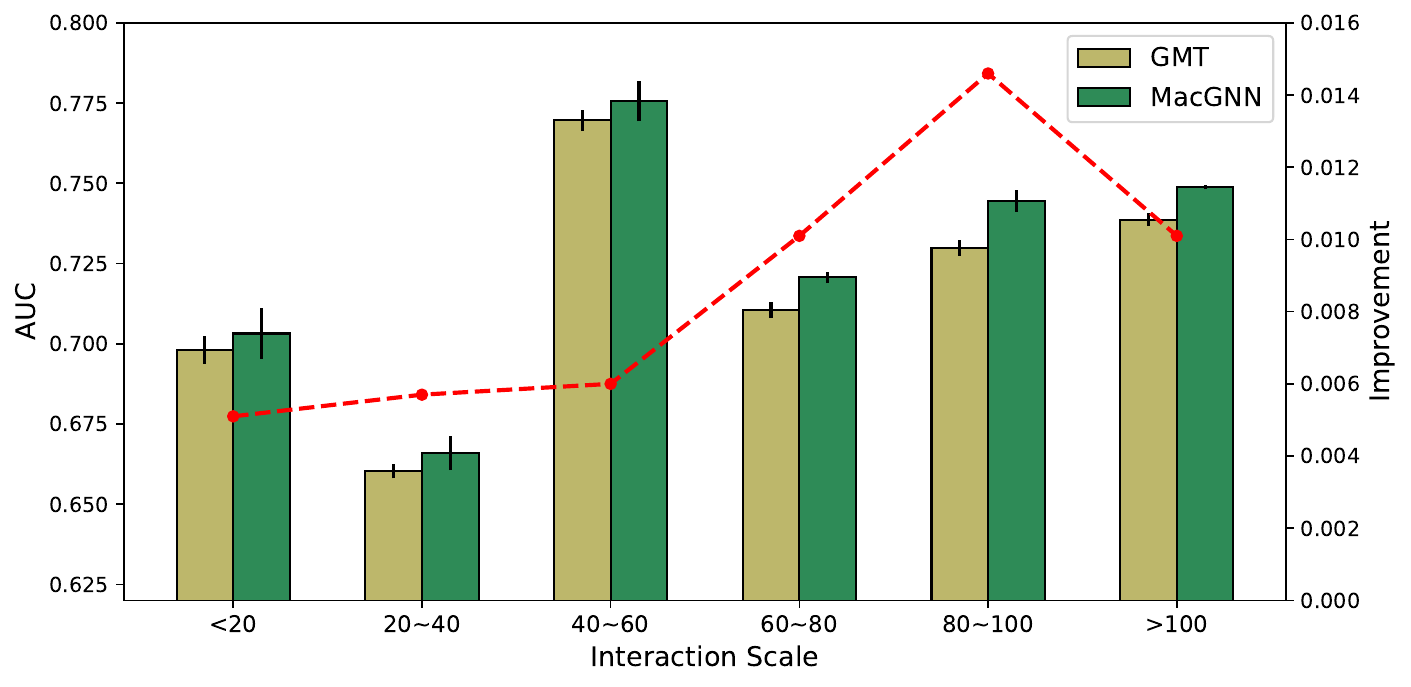}
    \caption{Case study of user groups with different interaction scales on the MovieLens dataset.}
    \label{fig:case_study}
\end{figure}

We further conduct the case study to verify the performance of \model on users with different interaction frequencies. Specifically, we divided users into 6 groups according to their interaction frequency on the MovieLens dataset. The case study results are illustrated in Figure \ref{fig:case_study}.


We can find that our \model performs better in most cases, which shows that the introduction of MAG can benefit users with different interaction frequencies.
This observation can be explained in the following two main aspects: 
(i) For low-active users, the modeling view from a macro perspective will bring additional general key features, and the high-order graph information from MAG also provides helpful information for user modeling. 
(ii) For high-active users, in addition to ensuring computational efficiency, macro modeling on MAG can also avoid noise and overly complex information contained in excessively long interaction sequences.
Thus, besides improving computational efficiency for considering both complete and high-order patterns, the organization of MAG is also beneficial for modeling interests in various interaction frequencies.

\end{document}